\documentclass[
 reprint,onecolumn,
 amsmath,amssymb,
 aps,
]{revtex4}

\usepackage{epsfig,bm,epsf,graphics}
\usepackage{graphicx}% Include figure files
\usepackage{dcolumn}% Align table columns on decimal point
\usepackage{bm}% bold math
\begin{document}
%\draft
\preprint{APS/123-QED}

\title{Theoretical methods for understanding advanced magnetic materials:
the case of frustrated thin films}% Force line breaks with \\

\author{H. T. Diep\footnote{diep@u-cergy.fr}}
\affiliation{%
Laboratoire de Physique Th\'eorique et Mod\'elisation,
Universit\'e de Cergy-Pontoise, CNRS, UMR 8089\\
2, Avenue Adolphe Chauvin, 95302 Cergy-Pontoise Cedex, France.\\
 }%


\begin{abstract}
Materials science has been intensively developed during the last 30 years. This is due, on the one hand, to an increasing demand of new materials for new applications and, on the other hand, to technological progress which allows for the synthesis of materials of desired characteristics and to investigate their properties with sophisticated experimental apparatus. Among these advanced materials, magnetic materials at nanometric  scale such as ultra thin films or ultra fine aggregates are no doubt among the most important for electronic devices.

In this review, we show advanced theoretical methods and solved examples that help understand microscopic mechanisms leading to experimental observations in magnetic thin films. Attention is paid to the case of magnetically frustrated systems in which two or more magnetic interactions are present and competing.  The interplay between spin frustration  and surface effects is the origin of spectacular phenomena which often occur at boundaries of phases with different symmetries: reentrance, disorder lines, coexistence of order and disorder at equilibrium.  These phenomena are shown and explained using of some exact methods, the Green's function and Monte Carlo simulation.  We show in particular how to calculate surface spin-wave modes, surface magnetization, surface reorientation transition and spin transport.
\vspace{0.5cm}
\begin{description}
\item[PACS numbers:  05.10.-a , 05.50.+q , 64.60.Cn , 75.10.-b , 75.10.Jm , 75.10.Hk , 75.70.-i ]
\end{description}
\end{abstract}

\pacs{Valid PACS appear here}% PACS, the Physics and Astronomy
\maketitle

%\tableofcontents
%\begin{keyword}
%Theory of Magnetism \sep Magnetic Thin Films \sep Surface Spin Waves \sep %Frustrated Spin Systems \sep Magnetic Materials \sep Phase Transition \sep %Monte Carlo Simulation \sep Statistical Physics
%\end{keyword}




\section{Introduction}
Material science has made a rapid and spectacular progress during the last 30 years, thanks to the advance of experimental investigation methods and a strong desire of scientific community to search for new and high-performance materials for new applications. In parallel to this intensive development, many efforts have been devoted to understanding theoretically microscopic mechanisms at the origin of the properties of new materials. Each kind of material needs specific theoretical
methods in spite of the fact that there is a large number of common basic principles that govern main properties of each material family.

In this paper, we confine our attention to the case of magnetic thin films. We would like to show basic physical principles that help us understand their general properties.  The main purpose of the paper is not to present technical details of each of them, but rather to show what can be understood using each of them. For technical details of a particular method, the reader is referred to numerous references given in the paper.  For demonstration purpose, we shall use magnetically frustrated thin films throughout the paper. These systems combine two difficult subjects: frustrated spin systems and surface physics. Frustrated spin systems have been subject of intensive studies during the last 30 years \cite{DiepFSS}.  Thanks to these efforts many points have been well understood in spite of the fact that there remains a large number of issues which are in debate. As seen below, frustrated spin systems contain many exotic properties such as high ground-state degeneracy, new symmetries, successive phase transitions, reentrant phase and disorder lines. Frustrated spin systems serve as ideal testing grounds for theories and approximations.  On the other hand, during the same period surface physics has also been widely investigated both experimentally and theoretically.  Thanks to technological progress, films and surfaces with desired properties could be fabricated and characterized with precision.  As a consequence, one has seen over the years numerous technological applications of thin films, coupled thin films and super-lattices, in various domains such as magnetic sensors, magnetic recording and data storage.  One of the spectacular effects is the colossal magnetoresistance \cite{Fert,Grunberg} which yields very interesting transport properties.  The search for new effects with new mechanisms in other kinds of materials continues intensively nowadays as never before.

Section \ref{main} is devoted to the presentation of the main theoretical background and concepts to understand frustrated spin systems and surface effects in magnetic materials. Needless to say, one cannot cover all recent developments in magnetic materials but an effort is made to outline the most important ones in our point of view. Section \ref{theory} is devoted to a few examples to illustrate striking effects due to the frustration and to the presence of a surface.
Concluding remarks are given in Section \ref{Conclu}.

\section{Background}\label{main}
\subsection{Theory of Phase Transition}
Many materials exhibit a phase transition.  There are several kinds of transition, each transition  is driven by the change of a physical parameter such as pressure, applied field, temperature ($T$), ... The most popular and most studied transition is no doubt the one corresponding to the passage  from a disordered phase to an ordered phase at the so-called magnetic ordering temperature or Curie temperature $T_c$. The transition is accompanied by a symmetry breaking. In general when the symmetry of one phase is a subgroup of the other phase the transition is continuous, namely the first derivatives of the free energy such as internal energy and magnetization are continuous functions of $T$. The second derivatives such as specific heat and susceptibility, on the other hand, diverge at $T_c$. The correlation length is infinite at $T_c$.  When the symmetry of one phase is not a symmetry subgroup of the other, the transition is in general of first order: the first derivatives of the free energy are discontinuous at $T_c$.  At the transition, the correlation length is finite and often there is a coexistence of the two phases.  For continuous transitions, also called second-order transition, the nature of the transition is characterized by a set of critical exponents which defines its "universality class".  Transitions in different systems may belong to the same universality class.

Why is the study of a phase transition interesting?  As the theory shows it, the characteristics of a transition are intimately connected to microscopic interactions between particles in the system.

The theory of phase transitions and critical phenomena has been intensively developed by Landau and co-workers since the 50's in the framework of the mean-field theory. Microscopic concepts have been introduced only in the early 70's with the renormalization group \cite{Wilson,Amit,Zinn}.  We have since then a clear picture of the transition mechanism and a clear identification of principal ingredients which determine the nature of the phase transition.  In fact, there is a small number of them such as the space dimension, the nature of interaction and the symmetry of the order parameter.

\subsection{Frustrated Spin Systems}
A spin is said "frustrated" when it cannot fully satisfy all the interactions with its neighbors. Let us take a triangle with an antiferromagnetic interaction $J(<0)$ between two sites: we see that we cannot arrange three Ising spins ($\pm1)$ to satisfy all three bonds. Among them, one spin satisfies one neighbor but not the other. It is frustrated. Note that any of the three spins can be in this situation. There are thus three equivalent configurations and three reverse configurations, making 6 the number of "degenerate states".  If we put  XY spins on such a triangle, the configuration with a minimum energy is the so-called "120-degree structure" where the two neighboring spins make a 120$^0$ angle. In this case, each interaction bond has an energy equal to $|J|\cos (2\pi/3)=-|J|/2$, namely half of the full interaction: the frustration is equally shared by three spins, unlike the Ising case.  Note that if we go from one spin to the neighboring spin in the trigonometric sense we can choose $ \cos (2\pi/3)$ or $-\cos (2\pi/3)$ for the turn angle: there is thus a two-fold degeneracy in the XY spin case. The left and right turn angles are called left and right chiralities. In an antiferromagnetic triangular lattice, one can construct the spin configuration from triangle to triangle.
The frustration in lattices with triangular plaquettes as unit such as in face-centered cubic and hexagonal-close-packed lattices is called "geometry frustration". Another category of frustration is when there is a competition between different kinds of incompatible interactions which results in a situation where no interaction is fully satisfied.  We take for example a square with three ferromagnetic bonds $J(>0)$ and one antiferromagnetic bond -$J$, we see that we cannot "fully" satisfy all bonds with Ising or XY spins put on the corners.

Frustrated spin systems are therefore very unstable systems with often very high ground-state degeneracy. In addition, novel symmetries can be induced such as the two-fold chirality seen above. Breaking this symmetry results in an Ising-like transition in a system of XY spins \cite{DiepBerge,DiepHassan}.  As will be seen in some examples below, the frustration is the origin of many spectacular effects such as  non collinear ground-state configurations, coexistence of order and disorder,  reentrance,  disorder lines, multiple phase transitions, etc.

\subsection{Surface Magnetism}

In thin films the lateral sizes are supposed to be infinite while the thickness is composed of a few dozens of atomic layers.
Spins at the two surfaces of a film lack a number of neighbors and as a consequence surfaces have physical properties different from the bulk. Of course, the difference is more pronounced if, in addition to the lack of neighbors, there are deviations of bulk parameters such as exchange interaction, spin-orbit coupling and magnetic anisotropy, and the presence of surface defects and impurities.  Such changes at the surface can lead to surface phase transition separated from the bulk transition, and surface reconstruction, namely change in lattice structure, lattice constant \cite{Bocchetti}, magnetic ordering, ... at the surface \cite{Heinrich,Zangwill,Binder-surf}.

Thin films of different materials, different geometries, different lattice structures, different thicknesses ... when  coupled give surprising results such as colossal magnetoresistance \cite{Fert,Grunberg}. Microscopic mechanisms leading to these striking effects are multiple.  Investigations on new artificial architectures for new applications are more and more intensive today. In the following section, we will give some basic microscopic mechanisms based on elementary excitations due to the film surface which allows for  understanding macroscopic behaviors of physical quantities such as surface magnetization, surface phase transition and  transition temperature.

\subsection{Methods}
To study properties of materials one uses various theories in condensed matter physics constructed from quantum mechanics and statistical physics \cite{Ashcroft,Chaikin}. Depending on the purpose of the investigation, we can choose many standard methods at hand (see details in Refs. \cite{DiepTM,DiepSP}):

(i) For a quick obtention of a phase diagram in the space of physical parameters such as temperature, interaction strengths, ... one can use a mean-field theory if the system is simple with no frustration, no disorder, ... Results are reasonable in three dimensions, though critical properties cannot be correctly obtained

(ii) For the nature of phase transitions and their criticality, the renormalization group  \cite{Wilson,Amit,Zinn} is no doubt the best tool. However for complicated systems such as frustrated systems, films and dots, this method is not easy to use

(iii) For low-dimensional systems with discrete spin models, exact methods can be used

(iv) For elementary excitations such as spin waves, one can use the classical or quantum spin-wave theory to get the spin-wave spectrum. The advantage of the spin-wave theory is that one can keep track of the microscopic effect of a given parameter on macroscopic properties of a magnetic system at low temperatures with a correct precision

(v) For quantum magnetic systems, the Green's function method allows one to calculate at ease the spin-wave spectrum, quantum fluctuations and thermodynamic properties up to rather high temperatures in magnetically ordered materials.  This method can be used for collinear spin states and non collinear (or canted) spin configurations as seen below

(vi) For all systems, in particular for complicated systems where analytical methods cannot be easily applied, Monte Carlo simulations can be used to calculate numerous physical properties, specially in the domain of phase transitions and critical phenomena as well as in the spin transport as seen below.

In the next section, we will show some of these methods and how they are practically applied to study various properties of thin films.

\section{Frustrated Thin Films}\label{theory}
\subsection{Exactly Solved Two-dimensional Models}
 Why are exactly solved models interesting? There are several reasons to study such models:
\begin{itemize}
\item Many hidden properties of a model cannot be revealed without exact mathematical demonstration
\item We do not know of any real material which corresponds to an exactly solved model, but we know that real materials should bear physical features which are not far from properties described in some exactly solved models if similar interactions are thought to exist in these materials.
\item Macroscopic effects observed in experiments cannot always allow us to find their origins if we do not already have some theoretical pictures provided by exact solutions in mind.
\end{itemize}

To date, only systems of discrete spins in one and two dimensions (2D) with short-range interactions can be exactly solved. Discrete spin models include Ising spin, $q$-state Potts models and some Potts clock-models.  The reader is referred to the book by Baxter \cite{Baxter} for principal exactly solved models.  In general, one-dimensional (1D) models with short-range interaction do not have a phase transition at a finite temperature. If infinite-range interactions are taken into account, then they have, though not exactly solved, a transition of second or first order depending on the decaying power of the interaction \cite{Bayong,Bayong2,Reynal1,Reynal2}. In 2D, most systems of discrete spins have a transition at a finite temperature. The most famous model is the 2D Ising model with the Onsager's solution \cite{Onsa}.

In this paper,  we are also interested in frustrated 2D systems because thin films in a sense are quasi two-dimensional.  We have exactly solved a number of frustrated Ising models such as  the Kagom\'e lattice \cite{Diep87K}, the generalized Kagom\'e lattice \cite{Diep91K}, the generalized honeycomb lattice \cite{Diep91HC} and various dilute centered square lattices \cite{Diep92DL,Diep89SQ,Diep92JMMM}.

For illustration, let us show the case of a Kagom\'e lattice with nearest-neighbor (NN) and next-nearest neighbor (NNN) interactions.  As seen below this Kagom\'e model possesses all interesting properties of the other frustrated models mentioned above.

\begin{figure}[htb]
\centering
\includegraphics[width=5cm]{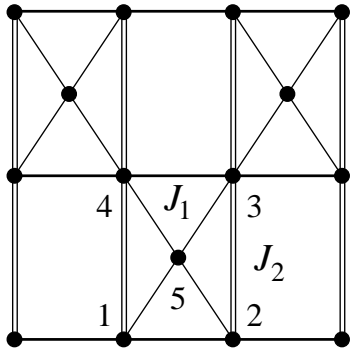}  % .eps
\caption{\label{re-fig7}
Kagom\'{e} lattice. Interactions between
nearest neighbors and between next-nearest neighbors, $J_{1}$ and
$J_{2}$, are shown by single and double bonds, respectively.  The
lattice sites in a cell are numbered for decimation demonstration.
}
\end{figure}

In general, 2D Ising models
without crossing interactions can be mapped onto the
16-vertex model or the 32-vertex model which satisfy the
free-fermion condition automatically as shown below with an Ising model defined on a Kagom\'{e}
lattice with interactions between NN
and between NNN, $J_{1}$ and $J_{2}$,
respectively, as shown in Fig. \ref{re-fig7}.
%Fig1

We consider the following Hamiltonian
\begin{equation}
H=-J_{1}\sum_{(ij)} \sigma_{i}\sigma_{j}-J_{2}\sum_{(ij)}
\sigma_{i}\sigma_{j}\label{re-eq19}
\end{equation}
%eq19
where $\sigma_i=\pm 1$ and the
first and second sums run over the spin pairs
connected by single and double bonds, respectively.  Note that the Kagom\'e original model, with antiferromagnetic $J_1$ and without $J_2$ interaction, has been exactly solved a long time ago showing no phase transition at finite temperatures \cite{Ka/Na}.

\begin{figure}[htb]
\centering
\includegraphics[width=8cm]{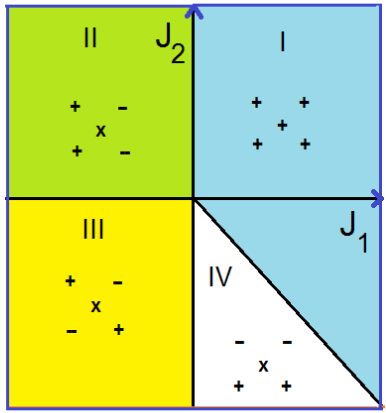}  % .eps
\caption{\label{re-gs}
Ground state of the Kagom\'{e} lattice in the space ($J_1,J_2)$. The spin configuration  is indicated in each of
the four zones I, II, III and IV: $\bf +$ for up spins, $\bf -$ for down spins, x for undetermined spins (free spins). The diagonal line separating zones I and IV is given by $J_2/J_1=-1$.
}
\end{figure}

The ground state (GS) of this model can be easily determined by an energy minimization. It is shown in Fig. \ref{re-gs} where one sees that only in zone I the GS is ferromagnetic. In other zones the central spin is undetermined because it has two up and two down neighbors, making its interaction energy zero: it is therefore free to flip. The GS spin configurations in these zones are thus "partially disordered".  Around the line $J_2/J_1=-1$ separating zone I and zone IV we will show below that many interesting effects occur when $T$ increases from zero.
%Fig2

\begin{figure}[htb]
\centering
\includegraphics[width=6cm]{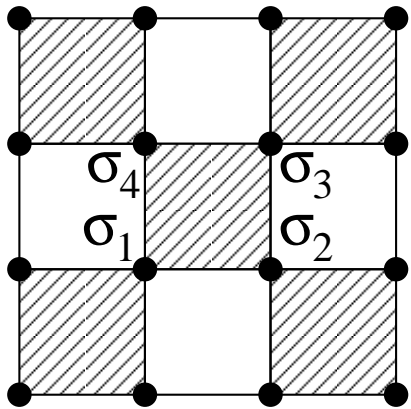}  % .eps
\caption{\label{re-fig8}
Checkerboard lattice. Each shaded
square is associated with the Boltzmann weight
$W(\sigma_{1},\sigma_{2},\sigma_{3},\sigma_{4})$, given in the
text.
}
\end{figure}

The partition function is written as

\begin{equation}
Z=\sum_{\sigma} \prod_{c} \exp[K_{1}(\sigma_{1}\sigma_{5}+
\sigma_{2}\sigma_{5}+\sigma_{3}\sigma_{5}+
\sigma_{4}\sigma_{5}+\sigma_{1}\sigma_{2}+
\sigma_{3}\sigma_{4})+K_{2}(\sigma_{1}\sigma_{4}+
\sigma_{3}\sigma_{2})]\label{eq20}
\end{equation}
%eq20
where $K_{1,2}=J_{1,2}/k_{B}T$  and where the sum is performed over all spin
configurations
and the product is taken over all elementary cells of the lattice.
To solve this model, we first
decimate the central spin of each elementary cell of the lattice and obtain a checkerboard Ising model
with multi-spin
interactions (see Fig. \ref{re-fig8}).
%Fig3

The Boltzmann weight of each shaded square is
given by

\begin{eqnarray}
W(\sigma_{1},\sigma_{2},
\sigma_{3},\sigma_{4})&= &2\cosh (K_{1}(\sigma_{1}+\sigma_{2}+
\sigma_{3}+\sigma_{4}))\exp[K_{2}(\sigma_{1}\sigma_{4}+
\sigma_{2}\sigma_{3}) \nonumber \\
 & &+K_{1}(\sigma_{1}\sigma_{2}+
\sigma_{3}\sigma_{4})]\label{eq21}
\end{eqnarray}
%eq21
The partition function of this checkerboard Ising model is
thus
\begin{equation}
Z=\sum_{\sigma} \prod W(\sigma_{1},\sigma_{2},
\sigma_{3},\sigma_{4})\label{eq22}
\end{equation}
%eq22
where the sum is performed over all spin configurations and
the product is taken over all the shaded squares of the
lattice.

\begin{figure}[htb]
\centering
\includegraphics[width=6cm]{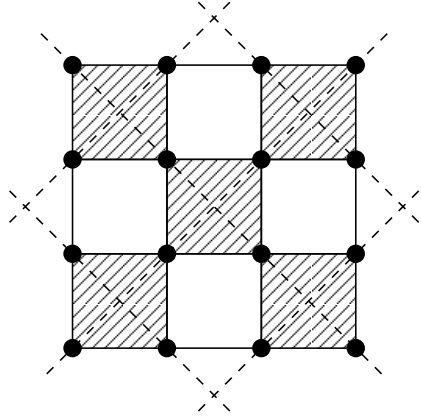}  % .eps
\caption{\label{re-fig9}
The checkerboard lattice and the
associated square lattice with their bonds indicated by dashed
lines.
}
\end{figure}

\begin{figure}[htb]
\centering
\includegraphics[width=6cm]{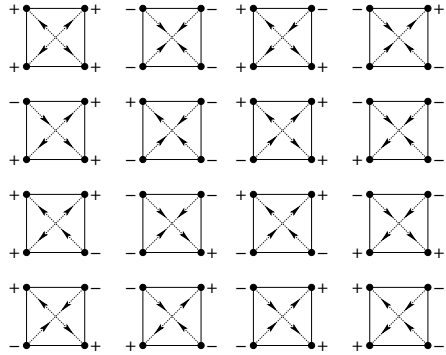}  % .eps
\caption{\label{re-fig10}
The relation between spin configurations
and arrow configurations of the associated vertex model.
}
\end{figure}

To map this model onto the 16-vertex model, we need to
introduce another square lattice where each site is placed at the
center of each shaded square of the checkerboard lattice, as shown
in Fig. \ref{re-fig9}.
%Fig4
At each bond of this lattice we associate an arrow pointing out of
the site if the Ising spin that is traversed by this bond is equal
to +1, and pointing into the site if the Ising spin is equal to
-1, as it is shown in Fig. \ref{re-fig10}.
In this way, we have a 16-vertex model on the associated
square lattice \cite{Baxter}. The Boltzmann weights of this vertex model
are expressed in terms of the Boltzmann weights of the
checkerboard Ising model, as follows
\begin{eqnarray}
\omega_{1}&= &W(-,-,+,+)\hspace{2.cm} \omega_{5}= W(-,+,-,+) \nonumber\\
\omega_{2}&= &W(+,+,-,-)\hspace{2.cm} \omega_{6}= W(+,-,+,-)\nonumber\\
\omega_{3}&= &W(-,+,+,-)\hspace{2.cm} \omega_{7}= W(+,+,+,+)\nonumber\\
\omega_{4}&= &W(+,-,-,+)\hspace{2.cm} \omega_{8}= W(-,-,-,-)\nonumber\\
\omega_{9}&= &W(-,+,+,+)\hspace{2.cm} \omega_{13}= W(+,-,+,+)\nonumber\\
\omega_{10}&= &W(+,-,-,-)\hspace{2.cm} \omega_{14}= W(-,+,-,-)\nonumber\\
\omega_{11}&= &W(+,+,-,+)\hspace{2.cm} \omega_{15}= W(+,+,+,-)\nonumber\\
\omega_{12}&= &W(-,-,+,-)\hspace{2.cm} \omega_{16}= W(-,-,-,+)\nonumber\\
& & \label{eq23}
\end{eqnarray}
%eq23
%\newpage
%Fig5
Taking Eq. (\ref{eq21}) into account, we obtain
\begin{eqnarray}
\omega_{1}&=&\omega_{2} = 2e^{-2K_{2}+2K_{1}}\nonumber\\
\omega_{3}&=&\omega_{4} = 2e^{2K_{2}-2K_{1}}\nonumber\\
\omega_{5}&=&\omega_{6} = 2e^{-2K_{2}-2K_{1}}\nonumber\\
\omega_{7}&=&\omega_{8} = 2e^{2K_{2}+2K_{1}}\cosh (4K_{1})\nonumber\\
\omega_{9}&=&\omega_{10} =\omega_{11}=\omega_{12} =
\omega_{13}=\omega_{14} = \omega_{15}=\omega_{16} =2\cosh
(2K_{1})\label{re-eq24}
\end{eqnarray}
%eq24

\begin{figure}[htb]
\centering
\includegraphics[width=8cm]{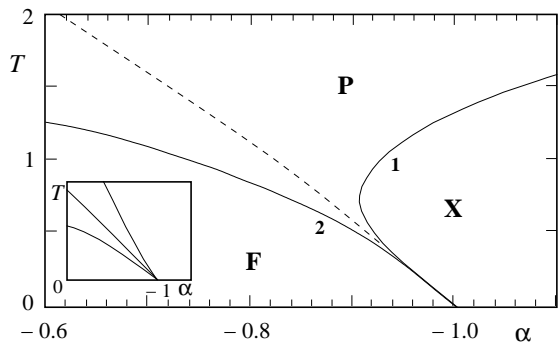}  % .eps
\caption{\label{re-fig20}
Phase diagram of the Kagom\'{e} lattice
with NNN interaction in the region $J_{1} > 0$ of the space
($\alpha=J_{2}/J_{1}, T$). $T$ is measured in the unit of
$J_{1}/k_{B}$.  Solid lines are critical lines, dashed line is the
disorder line. P, F and X stand for paramagnetic, ferromagnetic
and partially disordered phases, respectively.  The inset shows
schematically enlarged region of the endpoint.
}
\end{figure}

Generally, a vertex model is soluble if the vertex weights
satisfy the free-fermion conditions so that the partition function is
reducible to the $S$ matrix of a many-fermion system \cite {Gaff}. In
the present problem the free-fermion conditions  are the following
\begin{eqnarray}
\omega_{1}&= &\omega_{2}\:,\: \omega_{3}=\omega_{4}  \nonumber \\
\omega_{5}&= &\omega_{6} \:,\: \omega_{7}=\omega_{8}  \nonumber \\
\omega_{9}&= &\omega_{10}=\omega_{11}=\omega_{12}  \nonumber \\
\omega_{13}&= &\omega_{14}=\omega_{15}=\omega_{16}  \nonumber \\
\omega_{1}\omega_{3}&+&\omega_{5}\omega_{7}
-\omega_{9}\omega_{11}-\omega_{13}\omega_{15}=0\label{re-eq9}
\end{eqnarray}
As can be easily verified, Eqs.
(\ref{re-eq9}) are identically satisfied by the Boltzmann weights
Eqs. (\ref{re-eq24}), for arbitrary values of $K_{1}$ and $K_{2}$.
Using Eqs. (\ref{re-eq24}) for the 16-vertex model and calculating the
free energy of the model \cite{Diep87K,Baxter}
we obtain the critical condition for this system

\begin{eqnarray}
\frac {1}{2}\ [\exp(2K_{1}+2K_{2})\cosh (4K_{1})+
\exp(-2K_{1}-2K_{2})]&+& \nonumber\\
\cosh (2K_{1}-2K_{2})+2\cosh (2K_{1})=
2\mbox{max}\{\frac {1}{2}\ [\exp(2K_{1}&+&2K_{2})\cosh (4K_{1})+ \nonumber\\
\exp(-2K_{1}-2K_{2})]\:;\:\cosh (2K_{2}-2K_{1})&;&\cosh
(2K_{1})\}\label{eq25}
\end{eqnarray}
%eq25
This equation has up to four critical lines depending on
the values of $J_{1}$ and $J_{2}$.
For the
whole phase diagram, the reader is referred to Ref. \cite{Diep87K}.  We show in Fig. \ref{re-fig20} only the small
region of $J_{2}/J_{1}$ in the phase diagram which has two striking phenomena: the
reentrant paramagnetic phase and a disorder line.

The reentrant phase is defined as a paramagnetic phase which is located between two ordered phases on the temperature axis as seen in the region $-1<J_{2}/J_{1}<-0.91$: if we take for instance $J_{2}/J_{1}=-0.94$ and we go up on the temperature axis, we will pass through the ferromagnetic phase F, enter the "reentrant" paramagnetic phase, cross the disorder line, enter the partial disordered phase X where the central spins are free, and finally enter the paramagnetic phase  P \cite{Diep87K}.
The reentrant paramagnetic phase takes place thus between a low-$T$ ferromagnetic
phase and a partially disordered phase.
%Fig6

Note that in phase X, all central spins denoted by the number 5 in Fig. \ref{re-fig7} are free to flip at all $T$ while other spins are ordered up to the transition at
$T_{c}$. This
result shows an example where order and disorder  coexists in
an equilibrium state.

\begin{figure}[htb]
\centering
\includegraphics[width=5cm]{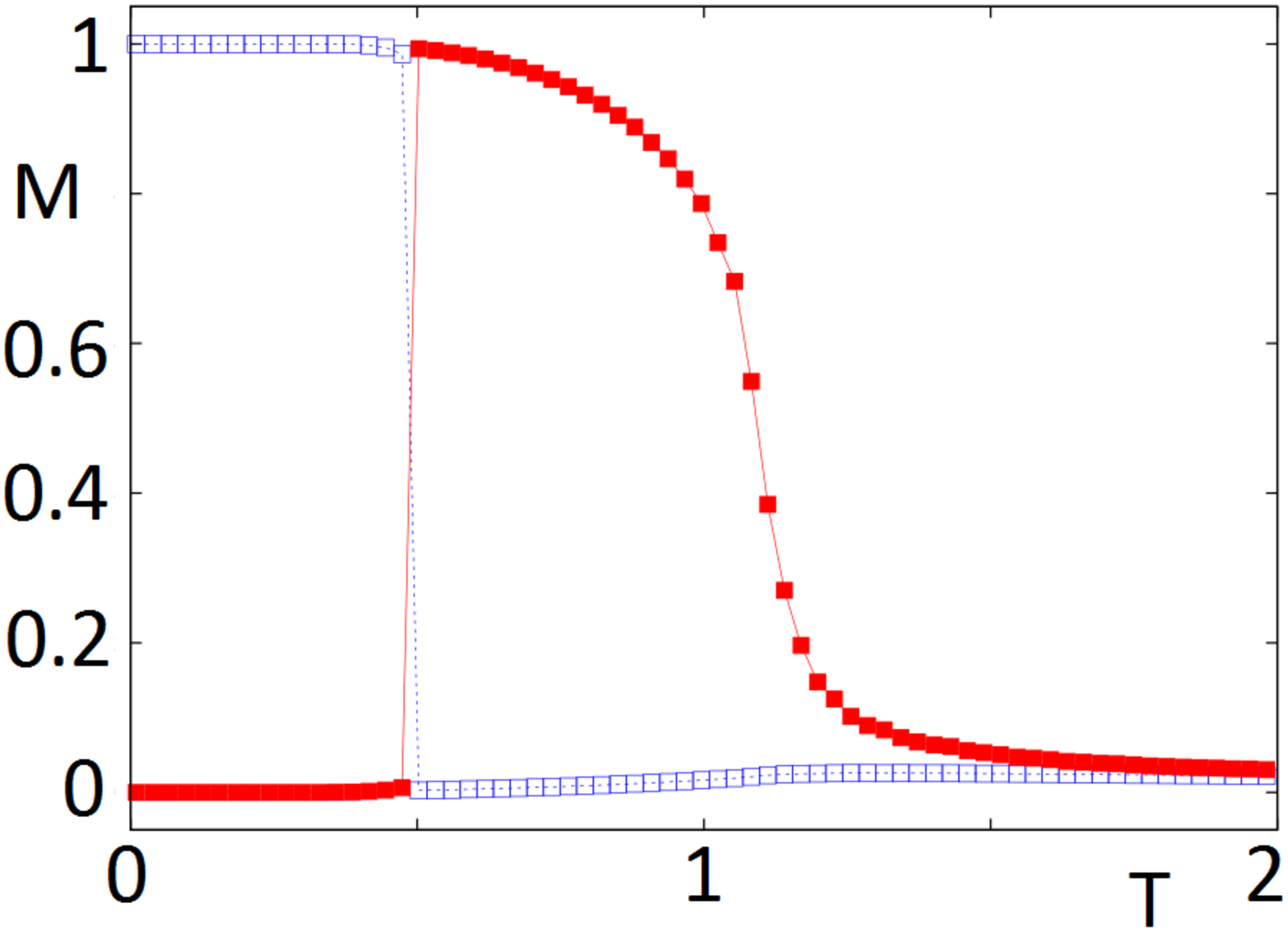}  % .eps
\includegraphics[width=5cm]{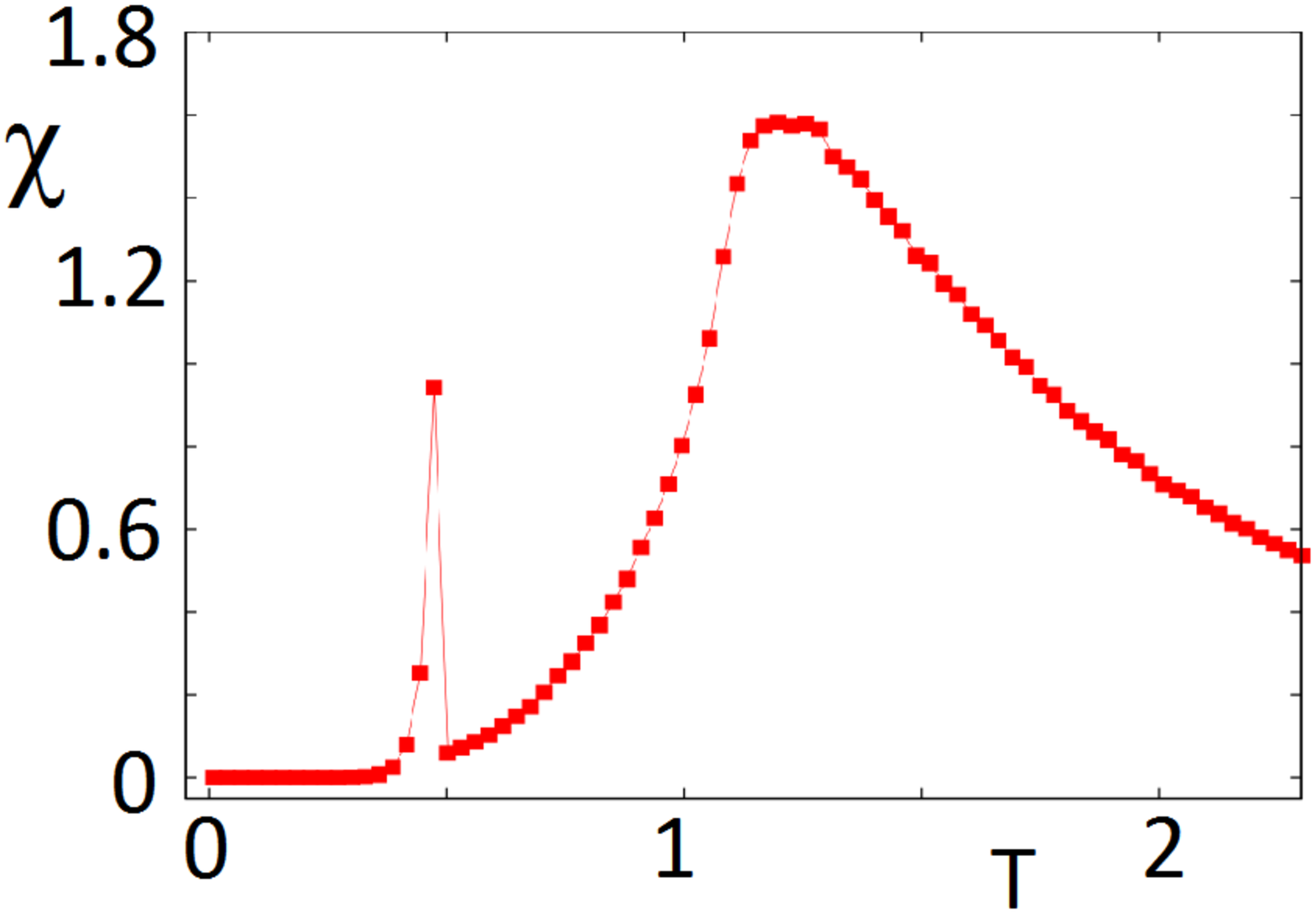}  % .eps
\caption{\label{re-kago}
Left: Magnetization of the sublattice 1 composed of cornered spins of ferromagnetic phase I (blue void squares) and the staggered magnetization defined for the phase IV of Fig. \ref{re-gs} (red filled squares) are shown in the reentrant region with $\alpha=J_{2}/J_{1}=-0.94$. See text for comments.
Right: Susceptibility of sublattice 1 versus $T$.
}
\end{figure}

It is important to note that though we get the exact solution for the critical surface, namely the exact location of the phase transition temperature in the space of parameters as shown in Fig. \ref{re-fig20}, we do not have the exact expression of the magnetization as a function of temperature. To verify the coexistence of order and disorder mentioned above we have to recourse to Monte Carlo simulations. This is easily done and the results for the order parameters and the susceptibility of one of them are shown in Fig. \ref{re-kago} for phases F and X at $J_{2}/J_{1}=-0.94$.   As seen, the F phase disappears at $T_1 \simeq 0.47$ and phase IV (defined in Fig. \ref{re-gs}) sets in at $T_2\simeq 0.50$ and disappears for $T>1.14$.  $T$ is measured in the unit of
$J_{1}/k_{B}$.  The paramagnetic region between $T_1$ and $T_2$ is the reentrant phase. Note that the disorder line discussed below cannot be seen by Monte Carlo simulations.
%Fig7

Let us now give the equation of the disorder line shown in Fig. \ref{re-fig20}:
\begin{equation}
e^{4K_{2}}=\frac {2(e^{4K_{1}}+1)}{e^{8K_{1}}+3}\label{eq43}
\end{equation}
%eq43
Usually, one defines each point on the disorder line as the temperature where
there is an effective reduction of dimensionality in such a way
that physical quantities become simplified spectacularly.
Along the disorder line, the partition
function is zero-dimensional and the correlation
functions behave as in one dimension (dimension reduction).
The disorder line is very important in
understanding the reentrance phenomenon. This type of line is necessary for the
change of ordering from the high-$T$ ordered phase to the low-$T$
one. In the narrow reentrant paramagnetic region, pre-ordering
fluctuations with different symmetries exist near each critical
line. Therefore the correlation functions  change their behavior
when crossing the "dividing line" as the temperature is varied in the reentrant paramagnetic region.
On this dividing line, or disorder line, the system "forgets" one dimension in order to adjust itself to the symmetry of the other side.
As a consequence of the change of symmetries there exist spins for
which the two-point correlation  function (between NN spins) has
different signs, near the two critical lines , in the reentrant
paramagnetic region. Hence it is reasonable to expect that it has
to vanish at a disorder temperature $T_{D}$. This point can be
considered as a non-critical transition point which separates two
different paramagnetic phases. The two-point correlation function
defined above may be thought of as a non-local "disorder
parameter". This particular point is just the one which has been
called a disorder point by Stephenson \cite{Ste} in analyzing the
behavior of correlation functions for systems with competing
interactions. Other models we solved have several disorder lines with dimension reduction \cite{Diep91HC,Diep92DL}
except the case of the centered square lattice where there is a disorder line without dimension reduction \cite{Diep89SQ}.

We believe that results of the exactly solved model in 2D shown above should also exist  in three dimensions (3D), though we cannot exactly solve 3D models.  To see this, we have studied a 3D version of the 2D Kagom\'e lattice which is a kind of body-centered lattice where the central spin in the lattice cell is free if the corner spins are in an antiferromagnetic order: the central spin has four up and four down neighbors making its energy zero as in the Kagom\'e lattice. We have shown that the partial disorder exists \cite{Quartu1997,Santa1} and the reentrant zone between phase F and phase X in Fig. \ref{re-fig20} closes up giving rise to a line of first-order transition \cite{Diep89BCC}.

To close this paragraph, we note that for other exactly solved frustrated models, the reader is referred to the review by Diep and Giacomini \cite{Diep-Giacomini}.

\subsection{Elementary Excitations: Surface Magnons}

We consider a thin film of $N_T$ layers with the Heisenberg quantum spin model.
The Hamiltonian is written as

\begin{eqnarray}
{\cal{H}}&=&-2\sum_{<i,j>}J_{ij} \vec S_i\cdot \vec S_j
-2\sum_{<i,j>}D_{ij} S_i^z S_j^z \nonumber\\
&=&  -2\sum_{\langle i,j\rangle}J_{ij} \left(
S_i^zS_j^z+\frac{1}{2}(S_i^+S_j^-+S_i^-S_j^+)\right)
-2\sum_{<i,j>} D_{ij}S_i^z S_j^z \nonumber\\
&& \label{surf30}
\end{eqnarray}
where $J_{ij}$ is positive (ferromagnetic) and $D_{ij}>0$ denotes  an exchange anisotropy.  When $D_{ij}$ is very large with respect to $J_{ij}$,  the spins have an Ising-like behavior.

For simplicity, let us suppose for the moment that all surface parameters are the same as the bulk ones with no defects and impurities. One of the microscopic mechanisms which govern thermodynamic properties of magnetic materials at low temperatures is the  spin waves. The presence of a surface often causes spin-wave modes localized at and near the surface. These modes cause in turn a diminution of the surface magnetization and the magnetic transition temperature. The methods to calculate the spin-wave spectrum from simple to more complicated are (see examples given in Ref. \cite{DiepTM}):

(i) the equation of motion written for spin operators $S_i^{\pm}$ of spin $\mathbf S_i$ occupying the lattice site $i$ of a given layer.  These operators are coupled to those of neighboring layers.  Writing an equation of motion for each layer, one obtains a system of coupled equations. Performing the Fourier transform in the $xy$ plane, one obtains the solution for the spin-wave spectrum.

(ii) the spin-wave theory using for example the Holstein-Primakoff spin operators for an expansion of the Hamiltonian. This is the second-quantization method. The harmonic spin-wave spectrum and nonlinear corrections can be obtained by diagonalizing the matrix written for operators of all layers.

(iii) the Green's function method using a correlation function between two spin operators. From this function one can deduce various thermodynamic quantities such as layer magnetizations and susceptibilities.  The advantage of this method is one can calculate properties up to rather high temperatures. However, with increasing temperature one looses the precision.

We summarize briefly here the principle of the Green's function method for illustration (see details in Ref. \cite{Diep1979,DiepTF91}).
We define one Green's function for each layer, numbering the surface as the first layer.
We write next the equation of motion for each of the Green's functions.
We obtain a system of coupled equations. We linearize these equations to reduce
higher-order Green's functions by using the Tyablikov decoupling scheme \cite{Tyablikov}.
We are then ready to make the Fourier transforms for all Green's functions in the $xy$ planes.
We obtain a system of equations in the space $(\vec k_{xy},\omega)$ where $\vec k _{xy}$ is
the wave vector parallel to the $xy$ plane and $\omega$ is the spin-wave frequency (pulsation).
Solving this system we obtain the Green's functions and $\omega$ as  functions of $\vec k _{xy}$.
Using the spectral theorem, we calculate the layer magnetization.  Concretely,
we define the following Green's function for two spins $\vec S_i$ and $\vec S_j$ as

\begin{equation}\label{gfd}
G_{i,j}(t,t')=\langle\langle S_i^+(t);S_j^-(t')\rangle\rangle
\end{equation}
The equation of motion of  $G_{i,j}(t,t')$ is written as

\begin{equation}\label{eqmvt}
i\hbar \frac{dG_{i,j}(t,t')}{dt}=(2\pi)^{-1}\langle[S_i^+(t),S_j^-(t')]\rangle +
\langle\langle [S_i^+;{\cal H}](t);S_j^-(t')\rangle\rangle
\end{equation}
\noindent where $[...]$ is the boson commutator and
$\langle...\rangle$ the thermal average in the canonical ensemble defined as
\begin{equation}
\langle F\rangle=\mbox{Tr}\mbox{e}^{-\beta {\cal{H}}}F/\mbox{Tr}\mbox{e}^{-\beta {\cal{H}}}
\end{equation}
\noindent with $\beta=1/k_BT$.
The commutator of the right-hand side of Eq. (\ref{eqmvt})
 generates functions of higher orders.  In a first
 approximation, we can reduce these functions with the help of the Tyablikov
 decoupling \cite{Tyablikov} as follows

\begin{equation}
\langle\langle S_m^zS_i^+;S_j^-\rangle\rangle
\simeq \langle S_m^z\rangle\langle\langle S_i^+;S_j^-\rangle\rangle,
\end{equation}
We obtain then the same kind of Green's function defined in Eq. (\ref{gfd}).
As the system is translation invariant in the $xy$ plane, we use the following Fourier transforms

\begin{equation}
G_{i,j}(t,t')=\frac{1}{\Delta} \int \int d{\vec k_{xy}} \frac{1}{2
\pi}\int^{+\infty}_{-\infty} d\omega\,\mbox{e}^{-i\omega( t-t')} \,
g_{n,n'}(\omega,\vec k_{xy})\, \mbox{e}^{i\vec k_{xy}.(\vec R_i-
\vec R_j)}
\end{equation}
where  $\omega$ is the  magnon pulsation (frequency), $\vec k_{xy}$ the wave vector
parallel to the surface,  $\vec R_i$ the position
of the spin at the site $i$, $n$ and $n'$ are respectively the indices of the planes to
which $i$ and $j$ belong ($n=1$ is the index of the surface).
The integration on $\vec k_{xy}$ is performed within the first Brillouin zone in the $xy$ plane.
Let  $\Delta$ be the surface of that zone.
Equation  (\ref{eqmvt}) becomes

\begin{equation}\label{greenf}
(\hbar\omega-A_n) g_{n,n'}+B_n(1-\delta_{n,1})g_{n-1,n'}+
C_n(1-\delta_{n,N_T}) g_{n+1,n'}= 2\delta_{n,n'}<S_n^z>
\end{equation}
where the factors $(1-\delta_{n,1})$ and $(1-\delta_{n,N_T})$
are added to ensure that there are no $C_n$ and $B_n$ terms for the first and the last layer.
The  coefficients $A_n$,  $B_n$ and  $C_n$ depend on the crystalline lattice of the film.
We give here some examples:

\begin{itemize}
\item Film of simple cubic lattice
\begin{eqnarray}
A_n&=& -2J_n<S_n^z>C\gamma_k+2C(J_n+D_n)<S_n^z> \nonumber\\
&&+2(J_{n,n+1}+D_{n,n+1})<S_{n+1}^z>\nonumber\\
&&+2(J_{n,n-1}+D_{n,n-1})<S_{n-1}^z> \\
B_n&=&2J_{n,n-1}<S_n^z> \\
C_n&=&2J_{n,n+1}<S_n^z>
\end{eqnarray}
where $C=4$ and $\gamma_k=\frac{1}{2}[\cos (k_xa)+\cos (k_ya)]$.

\item Film of body-centered cubic lattice
\begin{eqnarray}
A_n&=&8(J_{n,n+1}+D_{n,n+1})<S_{n+1}^z>\nonumber\\
&&+8(J_{n,n-1}+D_{n,n-1})<S_{n-1}^z> \\
B_n&=&8J_{n,n-1}<S_n^z>\gamma_k \\
C_n&=&8J_{n,n+1}<S_n^z>\gamma_k
\end{eqnarray}
where $\gamma_k=\cos (k_xa/2)\cos (k_ya/2)$
\end{itemize}

Writing Eq. (\ref{greenf}) for $n=1,2,...,N_T$, we obtain a system of $N_T$ equations which can be put in a matrix form

\begin{equation}\label{matrix}
{\bf M}(\omega){\bf g}={\bf u}
\end{equation}
where ${\bf u}$ is a column matrix whose n-th element is
$2\delta_{n,n'}<S_n^z>$.

For a given
$\vec k_{xy}$ the magnon dispersion relation
$\hbar\omega(\vec k_{xy})$ can be obtained by solving the secular equation $det |{\bf M}|=0$.
There are  $N_T$ eigenvalues $\hbar\omega_{i}$ ($i=1,...,N_T$)
for each $\vec k_{xy}$. It is obvious that $\omega_{i}$ depends
on all $\langle S_n^z\rangle$ contained in the coefficients $A_n$,
$B_n$ and $C_n$.

To calculate the thermal average of the  magnetization  of the layer
$n$  in the case where $S=\frac{1}{2}$, we use the following relation
(see chapter 6 of Ref.  \cite{DiepTM}):

\begin{equation}\label{lm}
\langle S_n^z\rangle=\frac{1}{2}-\langle S_n^-S_n^+\rangle
\end{equation}
where $\langle S_n^-S_n^+\rangle$ is given by the following spectral theorem

\begin{eqnarray}\label{fou}
   \langle {S^-_i} {S^+_j}\rangle &=&
   \lim_{\epsilon\to0}\frac{1}{\Delta}
   \int
   \int d{\vec k_{xy}}
   \int\limits_{-\infty}^{+\infty}\frac{i}{2\pi}
   \left[g_{n,n'}(\omega+i\epsilon)-
          g_{n,n'}(\omega-i\epsilon) \right] \nonumber\\
   &&\times\frac{d\omega}{e^{\beta\omega}-1}
   \mbox{e}^{i{\vec k_{xy}}.({\vec R_i}-{\vec R_j})}.
\end{eqnarray}
$\epsilon$ being an infinitesimal positive constant.
Equation  (\ref{lm}) becomes

\begin{equation}\label{lm1}
\langle S_n^z\rangle=\frac{1}{2}-
   \lim_{\epsilon\to0}\frac{1}{\Delta}
   \int
   \int d{\vec k_{xy}}
   \int\limits_{-\infty}^{+\infty}\frac{i}{2\pi}
   \left[ g_{n,n}(\omega+i\epsilon)-
          g_{n,n}(\omega-i\epsilon) \right]
\frac{d\omega}{\mbox{e}^{\beta\hbar \omega}-1}
\end{equation}
where the Green's function  $g_{n,n}$
is obtained by the solution of Eq. (\ref{matrix})

\begin{equation}\label{gnn}
g_{n,n}=\frac{|{\bf M}|_n}{|{\bf M}|}
\end{equation}
$|{\bf M}|_n $ is the determinant obtained by replacing
the n-th column of $|{\bf M}|$ by ${\bf u}$.

To simplify the notations we put $\hbar\omega_{i}=E_i$ and $\hbar\omega=E$
in the following.  By expressing
\begin{equation}
|{\bf M}|=\prod_i(E-E_{i})
\end{equation}
we see that $E_{i}$ ($i=1,...,N_T$) are the poles of the Green's function. We can therefore rewrite $g_{n,n}$ as

\begin{equation}\label{gnn1}
g_{n,n}=\sum_i\frac{f_n(E_i)}
{E-E_i}
\end{equation}
\noindent where $f_n(E_i)$ is given by
\begin{equation}\label{fn}
f_n(E_i)= \frac{|{\bf M}|_n(E_i)}
{\prod_{j\neq i}(E_i-E_{j})}
\end{equation}
Replacing Eq. (\ref{gnn1}) in Eq. (\ref{lm1}) and making use of the following identity

\begin{equation}\label{id}
\frac {1}{x-i\eta} - \frac {1}{x+i\eta}=2\pi i\delta (x)
\end{equation}
we obtain
\begin{equation}\label{lm2}
\langle S_n^z\rangle=\frac{1}{2}-
   \frac{1}{\Delta}
   \int
   \int dk_xdk_y
   \sum_{i=1}^{N_T}\frac{f_n(E_i)}
   {\mbox{e}^{\beta E_i}-1}
\end{equation}
where $n=1,...,N_T$.

As $<S_n^z>$ depends on the magnetizations of the neighboring layers via $E_i (i=1,...,N_T)$,
we should solve by iteration the equations
(\ref{lm2}) written for all layers, namely for  $n=1,...,N_T$, to obtain the layer magnetizations at a given temperature $T$.

The critical temperature $T_c$ can be calculated in a self-consistent manner by iteration, letting all  $<S_n^z>$  tend to zero.

\begin{figure}[htb]
\centering
\includegraphics[width=5cm]{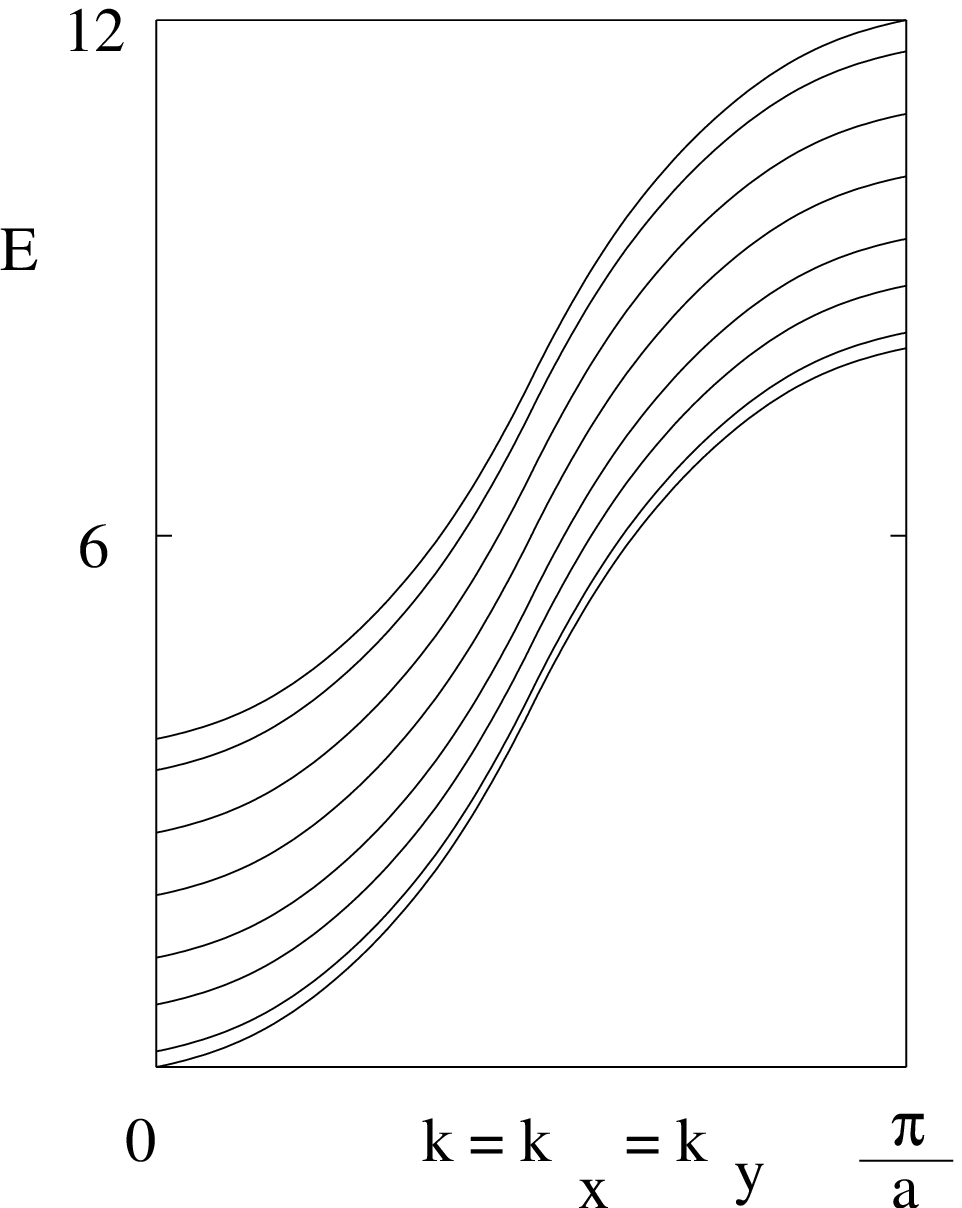}  % .eps
\includegraphics[width=5cm]{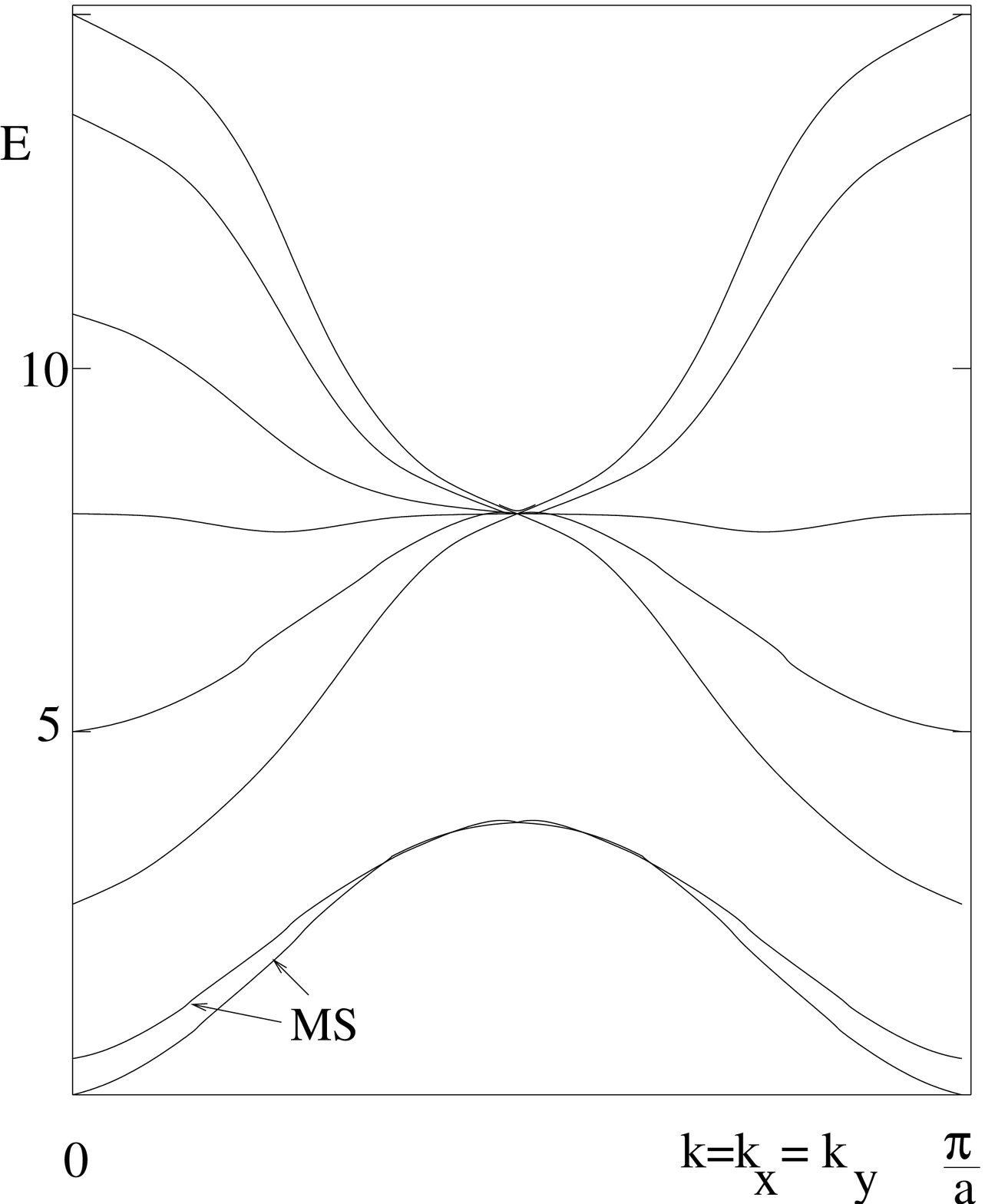}  % .eps
\caption{\label{ffig16_6}
Left: Magnon spectrum $E=\hbar\omega$ of a ferromagnetic film with a simple cubic lattice versus
$k\equiv k_x=k_y$ for $N_T=8$ and $D/J=0.01$.  No surface mode is observed for this case. Right: Magnon spectrum $E=\hbar\omega$ of a ferromagnetic film with a body-centered cubic lattice versus
$k\equiv k_x=k_y$ for $N_T=8$ and $D/J=0.01$. The branches of surface modes are indicated by  MS.
}
\end{figure}

Let us show in Fig. \ref{ffig16_6} two examples of spin-wave spectrum, one without surface modes as in a simple cubic film and the other with surface localized modes as in body-centered cubic ferromagnetic case.
%Fig8

It is very important to note that acoustic surface localized spin waves lie below the bulk frequencies so that these low-lying energies will give larger integrands to the integral on the right-hand side of Eq. (\ref{lm2}), making $<S_n^z>$ to be smaller. The same effect explains the diminution of $T_c$ in thin films whenever low-lying surface spin waves exist in the spectrum.

%\begin{figure}[htb]
%\centering
%\includegraphics[width=5cm]{fig16_7}  % .eps
%\caption{\label{ffig16_7}
%Magnon spectrum $E=\hbar\omega$ of a ferromagnetic film with a body-centered cubic lattice versus
%$k\equiv k_x=k_y$ for $N_T=8$ and $D/J=0.01$. The branches of surface modes are indicated by  MS.
%}
%\end{figure}

\begin{figure}[htb]
\centering
\includegraphics[width=10cm]{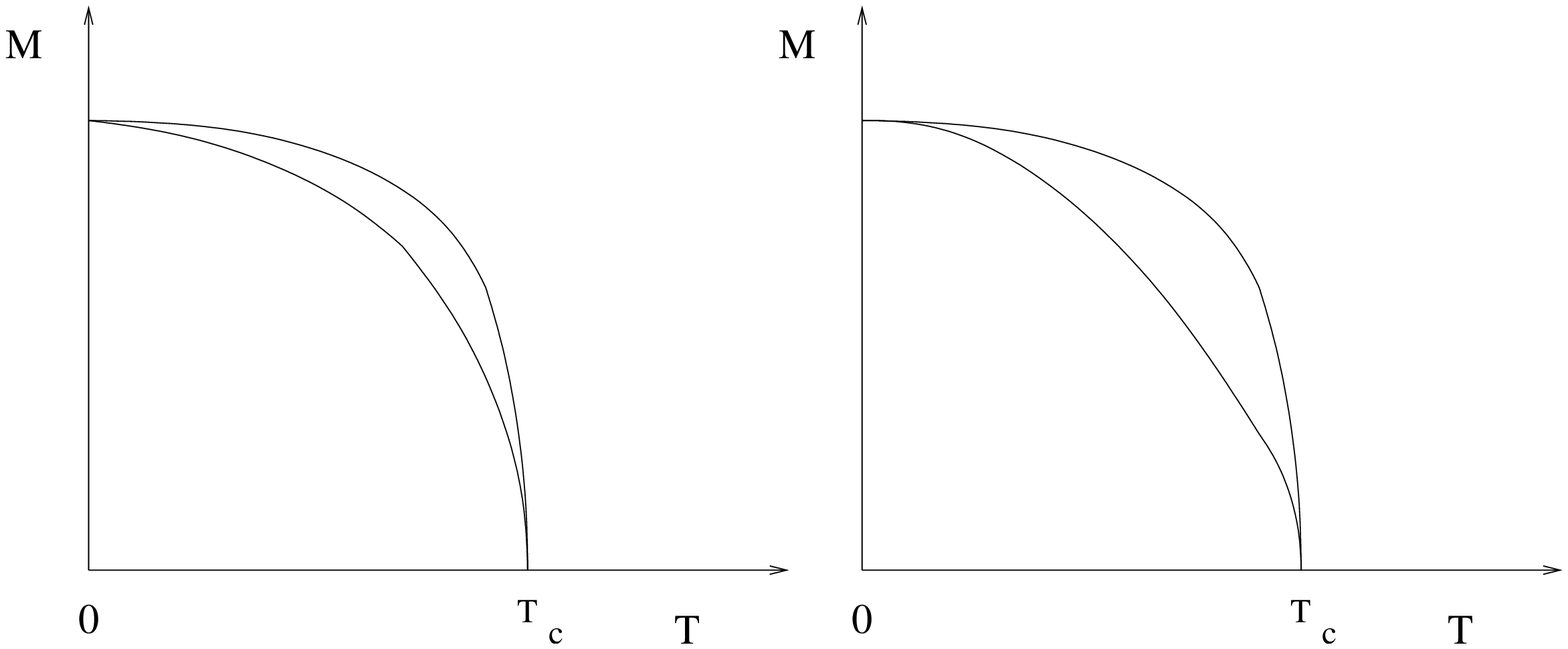}  % .eps
\caption{\label{ffig16_8}
Ferromagnetic films of simple cubic lattice (left) and body-centered cubic lattice (right): magnetizations of the surface layer
 (lower curve) and  the second layer (upper curve), with $N_T=4$, $D=0.01J$, $J=1$.
}
\end{figure}

Figure \ref{ffig16_8}  shows the results of the layer magnetizations for the first two layers in the films considered above with  $N_T=4$.
%Fig9

Calculations for antiferromagnetic thin films with collinear spin configurations can be performed using Green's functions \cite{Diep1979}.  The physics is similar with strong effects of localized surface spin waves and a non-uniform spin contractions near the surface at zero temperature due to quantum fluctuations \cite{DiepTF91}.

\subsection{Frustrated Films}\label{FTF}

We showed above for a pedagogical purpose a detailed technique for using the Green's function method. In the case of frustrated thin films, the ground-state spin configurations are not only non collinear but also non uniform from the surface to the interior layers.  In a class of helimagnets, the angle between neighboring spins is due to the competition between the NN and the NNN interactions.  Bulk spin configurations of such helimagnets were discovered more than 50 years ago by Yoshimori \cite{Yoshimori} and Villain \cite{Villain59}. Some works have been done to investigate the low-temperature spin-wave behaviors \cite{Harada,Rastelli,Diep89} and the phase transition \cite{Diep89b} in the bulk crystals.

For surface effects in frustrated films, a number of our works have been recently done among which we can mention the case of a frustrated surface on a ferromagnetic substrate film \cite{NgoSurface}, the fully frustrated antiferromagnetic face-centered cubic film \cite{NgoSurface2}, and very recently the helimagnetic thin films in zero field \cite{Diep2015,Sahbi} and under an applied field \cite{SahbiField}.

The Green's function method for non collinear magnets has been developed for the bulk crystal \cite{Quartu1998}.  We have extended this to the case of non collinear thin films in the works just mentioned. Since two spins $\mathbf S_i$ and $\mathbf S_j$ form an angle $\cos \theta_{ij}$ one can express the Hamiltonian in the local coordinates as follows \cite{Diep2015}:

\begin{eqnarray}
\mathcal H &=& - \sum_{<i,j>}
J_{i,j}\Bigg\{\frac{1}{4}\left(\cos\theta_{ij} -1\right)
\left(S^+_iS^+_j +S^-_iS^-_j\right)\nonumber\\
&+& \frac{1}{4}\left(\cos\theta_{ij} +1\right) \left(S^+_iS^-_j
+S^-_iS^+_j\right)\nonumber\\
&+&\frac{1}{2}\sin\theta_{ij}\left(S^+_i +S^-_i\right)S^z_j
-\frac{1}{2}\sin\theta_{ij}S^z_i\left(S^+_j
+S^-_j\right)\nonumber\\
&+&\cos\theta_{ij}S^z_iS^z_j\Bigg\}- \sum_{<i,j>}I_{i,j}S^z_iS^z_j\cos \theta_{ij}
\label{eq:HGH2}
\end{eqnarray}
The last term is an anisotropy added to facilitate a numerical convergence for  ultra thin films at long-wave lengths since it is known that in 2D there is no ordering for isotropic Heisenberg spins at finite temperatures \cite{Mermin}.

The determination of the angles in the ground state can be done either by minimizing the interaction energy with respect to interaction parameters or by the so-called steepest descent method which has been proved to be very efficient \cite{NgoSurface,NgoSurface2}.  Using their values, one can follow the different steps presented above for the collinear magnetic films, one then obtains a matrix which can be numerically diagonalized  to get the spin-wave spectrum which is used in turn to calculate physical properties in the same manner as for the collinear case presented above.

\begin{figure}[htb]
\centering
\includegraphics[width=5cm]{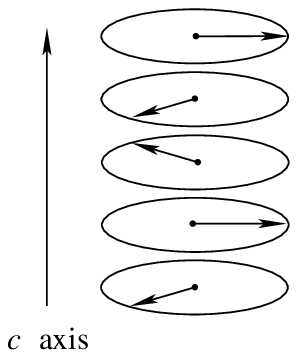}  % .eps
\includegraphics[width=6cm,angle=0]{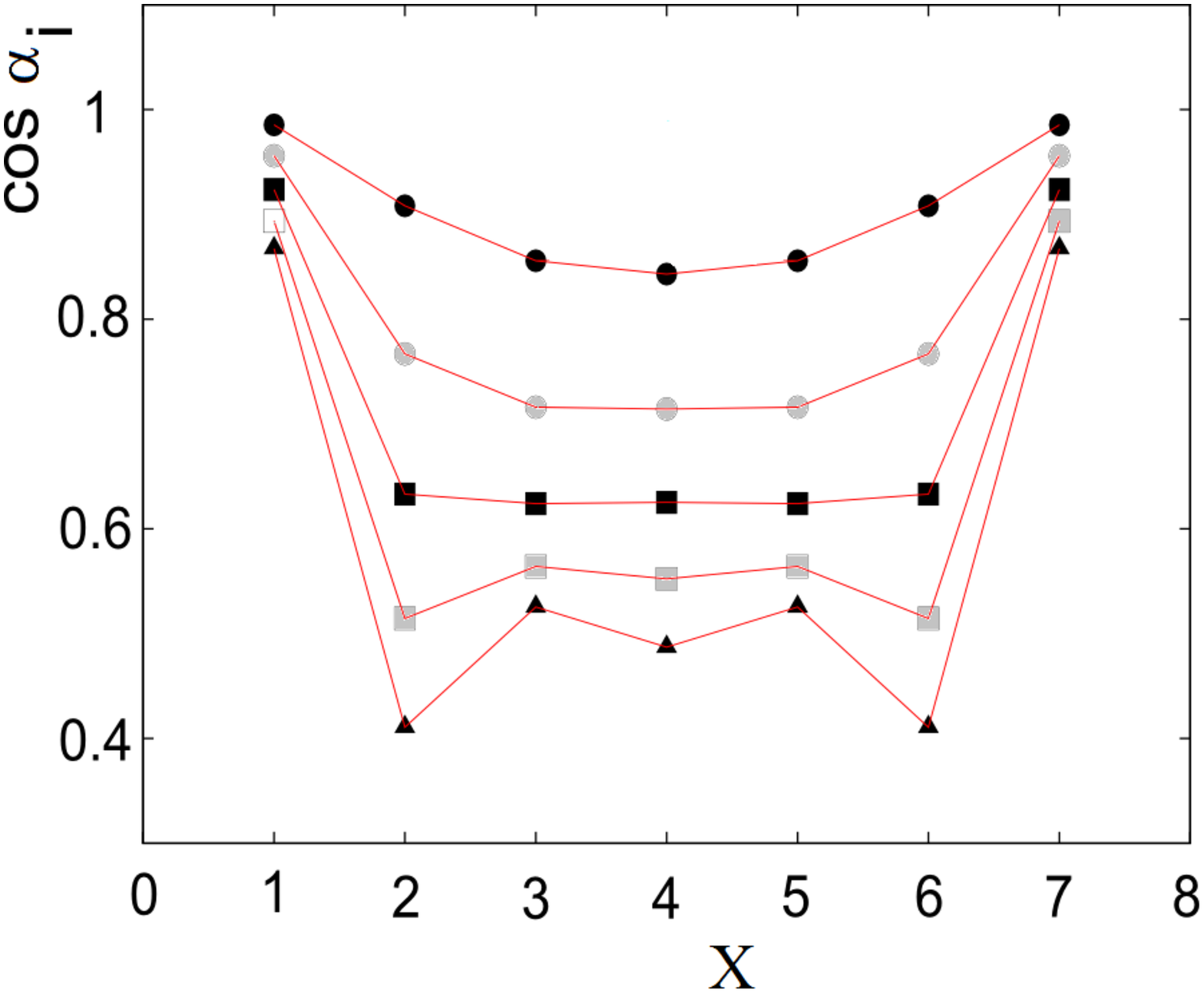}  % .eps
\caption{ Left: Bulk helical structure along the $c$-axis,
in the case $\alpha=2\pi/3$, namely $J_2/J_1=-2$.  Right: (color online) Cosinus of $\alpha_1=\theta_1-\theta_2$, ..., $\alpha_7=\theta_7-\theta_8$ across the film
for $J_2/J_1=-1.2,-1.4,-1.6,-1.8, -2$ (from top) with $N_z=8$: $\alpha_i$ stands for $\theta_i-\theta_{i+1}$ and $X$ indicates the film layer $i$ where the angle $\alpha_i$ with the layer $(i+1)$ is shown.  The values of the angles are given in Table 1: a strong rearrangement of spins near the surface is observed. }\label{GSA}
\end{figure}

Let us show the case of a helimagnetic film. In the bulk, the turn angle in one direction is determined by the ratio between the antiferromagnetic NNN interaction $J_2 (<0)$ and the NN interaction $J_1$.  For the body-centered cubic lattice, one has $\cos \theta=-J_1/J_2$.  The helimagnetic phase is stable for $|J_2|/J_1>1$.
Consider a film with the $c$ axis perpendicular to the film surface. For simplicity, one supposes the turn angle along the $c$ axis is due to  $J_2$. Because of the lack of neighbors,  the spins on the surface and on the second layer have the turn angles strongly deviated from the bulk value \cite{Diep2015}.
The results calculated for various $J_2/J_1$ are shown in Fig. \ref{GSA} (right) for a film of $N_z=8$ layers.
The values obtained are shown in Table \ref{table} where one sees that the angles near the surface (2nd and 3rd columns) are very different from that of the bulk (last column).

%\begin{widetext}
\begin{center}
\begin{table}
\begin{small}
\begin{tabular}{|l|c|c|c|c|r|}
\hline
$J_2/J_1$ & $\cos \theta_{1,2}$ & $\cos\theta_{2,3}$ & $\cos\theta_{3,4}$ & $\cos\theta_{4,5}$ & $\alpha$(bulk)  \\
\hline
&&&&&\\
-1.2  &  0.985($9.79^\circ$) &  0.908($24.73^\circ$)       &    0.855($31.15^\circ$)    &   0.843($32.54^\circ$)  &   $33.56^\circ$    \\
-1.4 &  0.955($17.07^\circ$)&  0.767($39.92^\circ$)  &  0.716($44.28^\circ$)    &   0.714($44.41^\circ$)   &  $44.42^\circ$     \\
-1.6 & 0.924($22.52^\circ$) &  0.633($50.73^\circ$) & 0.624($51.38^\circ$)  &  0.625($51.30^\circ$)   &  $51.32^\circ$        \\
-1.8     &  0.894($26.66^\circ$)  &  0.514($59.04^\circ$)  & 0.564($55.66^\circ$)   &  0.552($56.48^\circ$)  &  $56.25^\circ$    \\
-2.0       &  0.867($29.84^\circ$)  &  0.411($65.76^\circ$) &  0.525($58.31^\circ$)   &  0.487($60.85^\circ$)  & $60^\circ$   \\
&&&&&\\
\hline
\end{tabular}
\end{small}
\caption{Values of $\cos \theta_{n,n+1}=\alpha_n$ between two adjacent layers are shown for various
values of $J_2/J_1$. Only angles of the first half of the 8-layer film are shown: other angles are, by symmetry,
$\cos\theta_{7,8}$=$\cos\theta_{1,2}$, $\cos\theta_{6,7}$=$\cos\theta_{2,3}$, $\cos\theta_{5,6}$=$\cos\theta_{3,4}$. The values in parentheses are angles in degrees.
The last column shows the value of the angle in the bulk case (infinite thickness).
For presentation, angles are shown with two  digits. \label{table} }
\end{table}
\end{center}
%\end{widetext}

\vspace{1cm}
%Fig10

\begin{figure}[htb]
\centering
\includegraphics[width=6cm,angle=0]{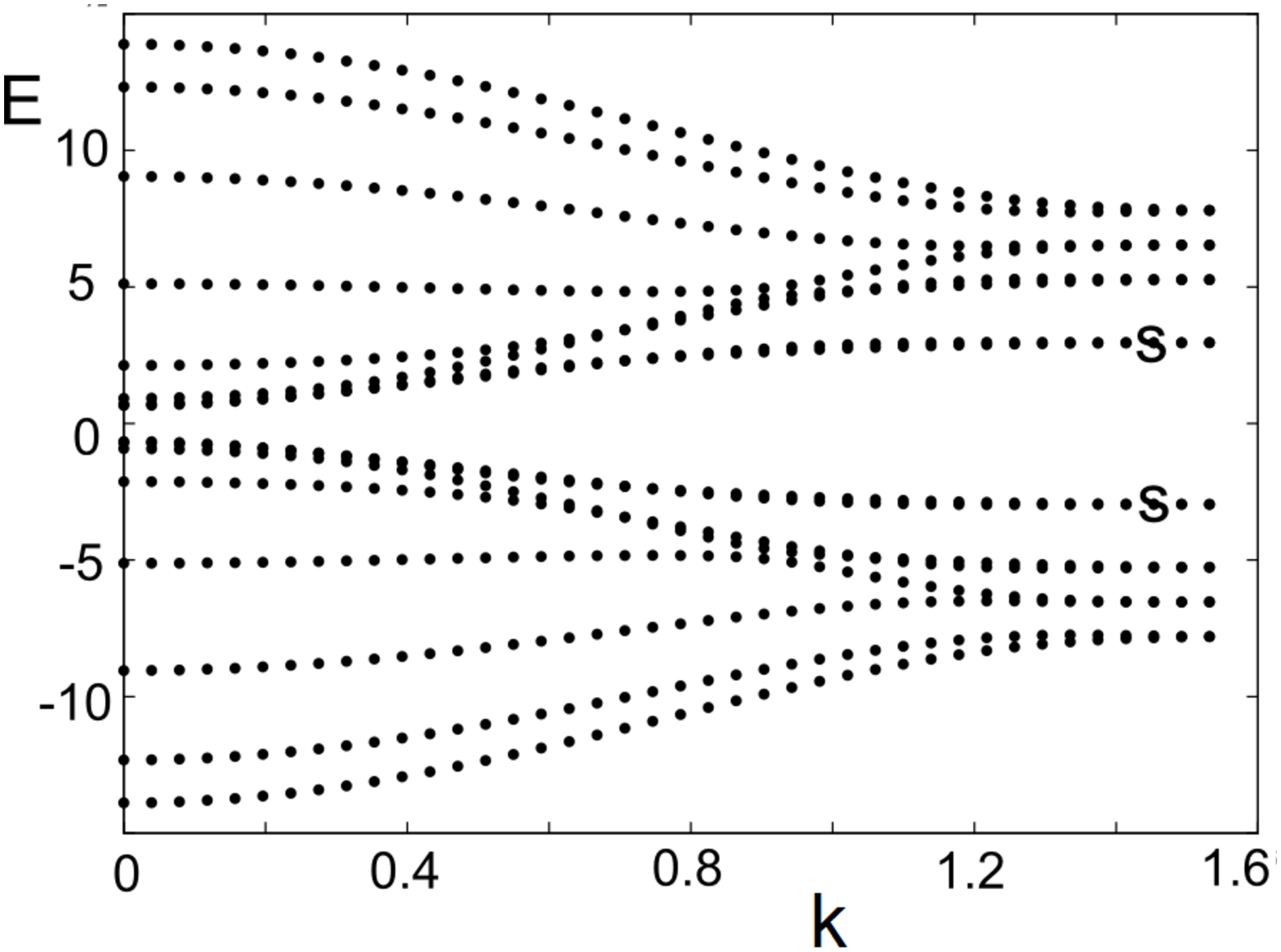}  % .eps
\includegraphics[width=6cm,angle=0]{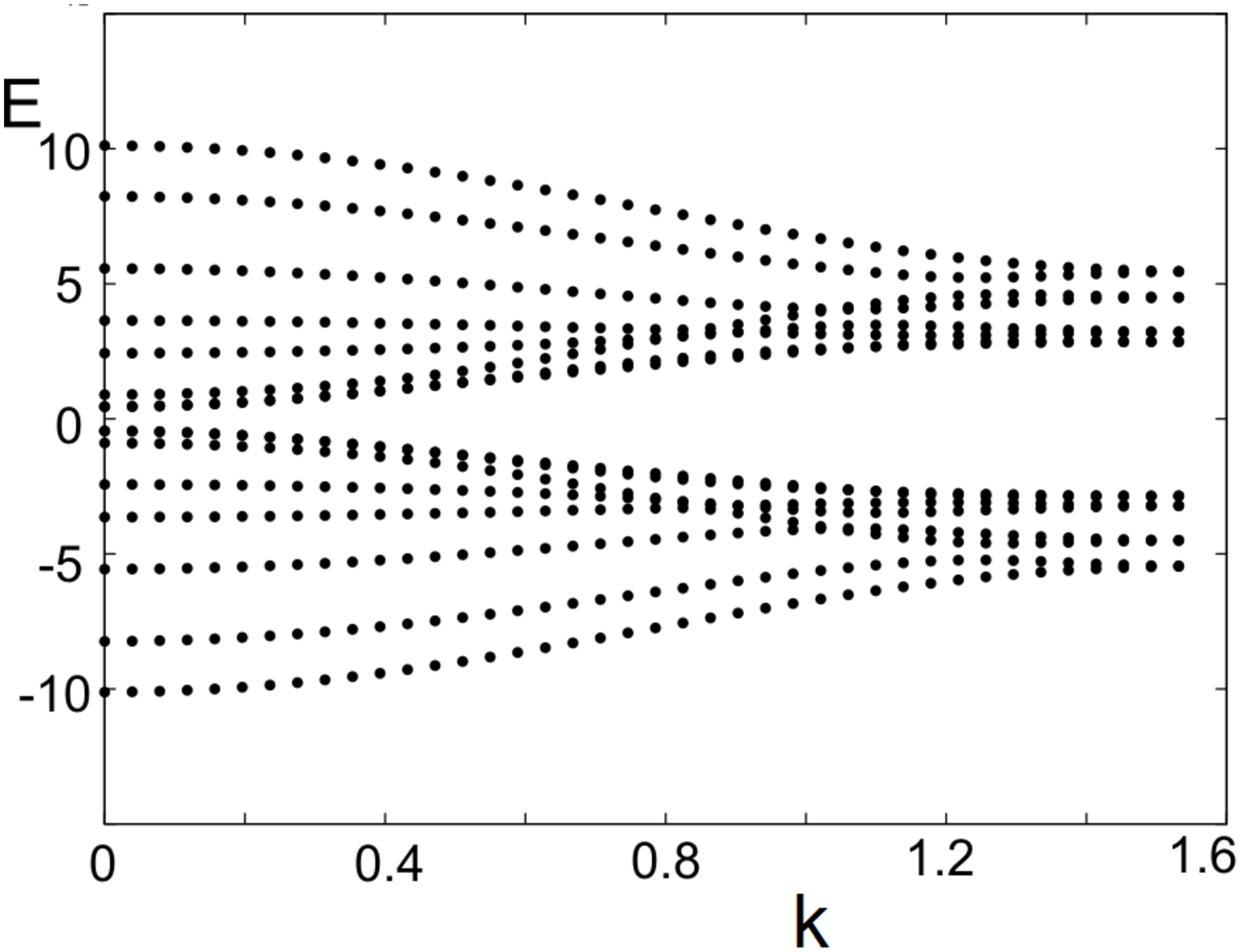}  % .eps
\caption{ Spectrum $E=\hbar \omega$ versus $k\equiv k_x=k_y$ for $J_2/J_1=-1.4$ at $T=0.1$ (left) and $T=1.02$ (right) for $N_z=8$ and $d=I/J_1=0.1$. The surface branches are indicated by $s$.}\label{sweta}
\end{figure}

\begin{figure}[htb]
\centering
\includegraphics[width=6cm,angle=0]{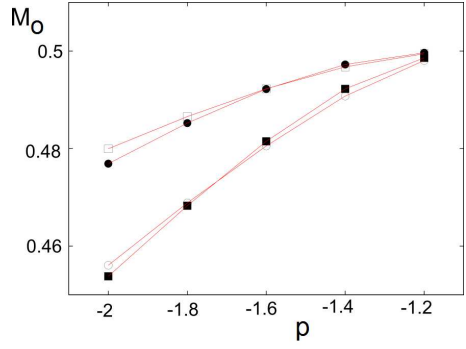}  % .eps
\caption{(Color online) Spin lengths of the first four layers at $T=0$ for several values of $p=J_2/J_1$ with $d=0.1$, $N_z=8$.
Black circles,  void circles, black squares and void squares are for first, second, third and fourth layers, respectively. See text for comments. }\label{spin0}
\end{figure}

The spectrum at two temperatures is shown in Fig. \ref{sweta} where the surface spin waves are indicated. The spin lengths at $T=0$ of the different layers are shown in Fig. \ref{spin0} as  functions of $J_2/J_1$.  When $J_2$ tends to -1, the spin configuration becomes ferromagnetic, the spin has the full length 1/2.
%Fig11

%Fig12

\begin{figure}[htb]
\centering
\includegraphics[width=6cm,angle=0]{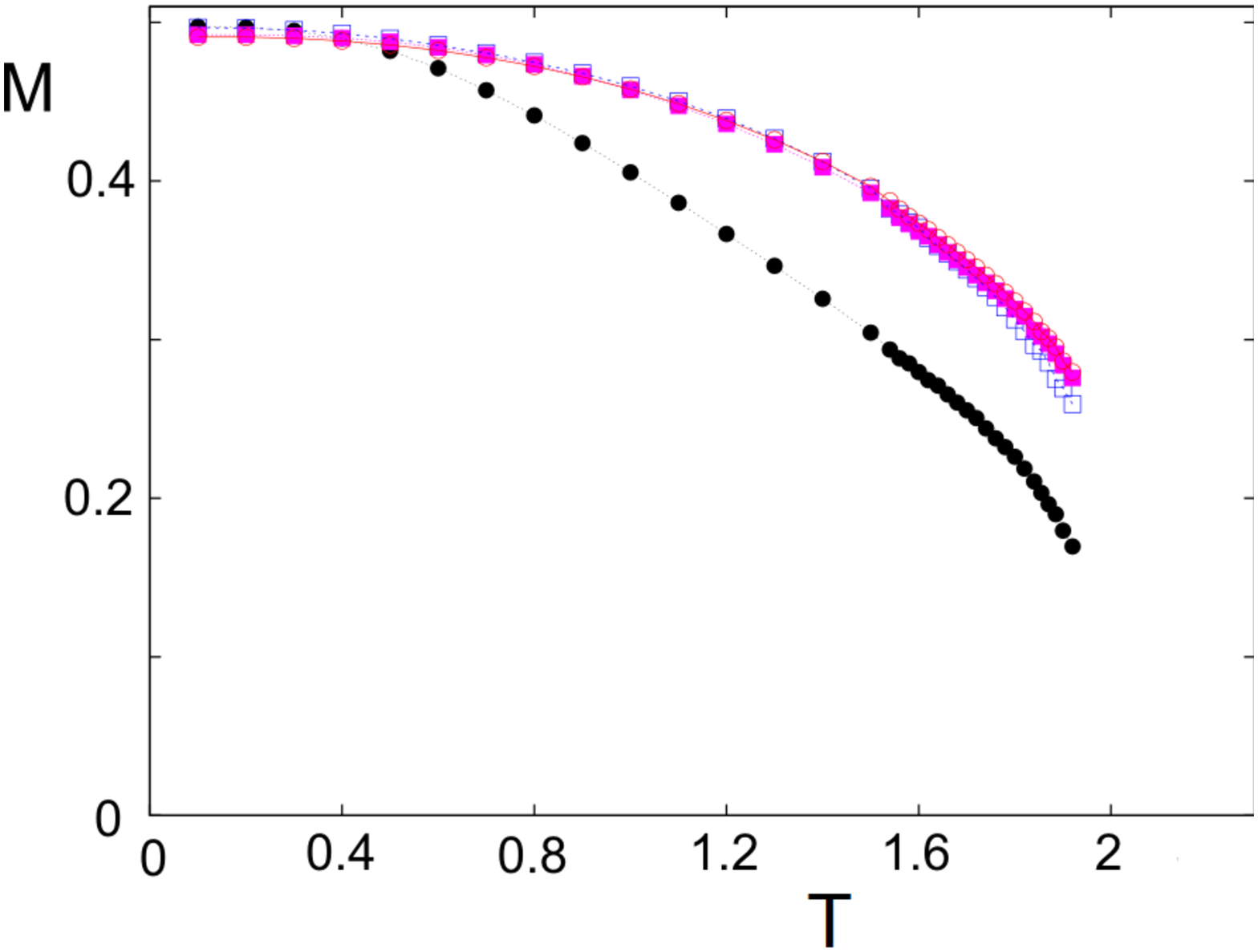}  % .eps
\includegraphics[width=6cm,angle=0]{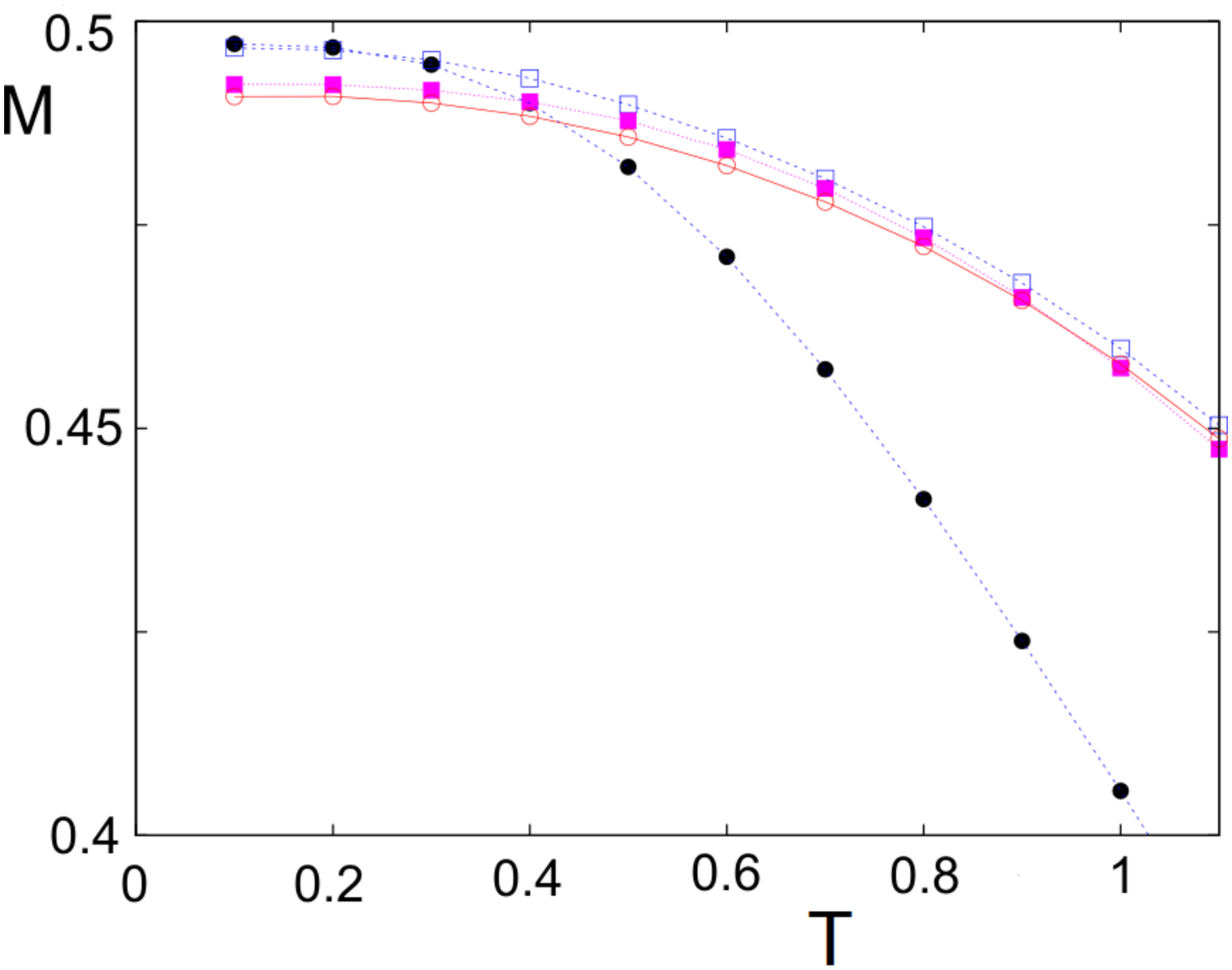}  % .eps
\caption{(Color online) Layer magnetizations as functions of $T$ for $J_2/J_1=-1.4$ with $d=0.1$, $N_z=8$ (left). Zoom of the region at low $T$ to show crossover (right). Black circles, blue void squares, magenta squares and red void circles are for first, second, third and fourth layers, respectively.  See text.}\label{magnet14}
\end{figure}

The layer magnetizations are shown in Fig. \ref{magnet14} where one notices the crossovers between them at low $T$.  This is due to the competition between quantum fluctuations, which depends on the strength of antiferromagnetic interaction, and the thermal fluctuations which depends on the local coordinations.

%Fig13

\subsection{Surface Disordering and Surface Criticality: Monte Carlo Simulations}
As said earlier, Monte Carlo methods can be used in complicated systems where analytical methods cannot be efficiently used.  Depending on the difficulty of the investigation, we should choose a suitable Monte Carlo technique.  For a simple investigation to have a rough idea about physical properties of a given system, the standard Metropolis algorithm is sufficient \cite{Metropolis,Binder}. It consists in calculating the energy  $E_1$ of a spin, then changing its state and calculating its new energy $E_2$. If $E_2<E_1$ then the new state is accepted. If $E_2>E_1$ the new state is accepted with a probability proportional to
$\exp[-(E_2-E_1)/(k_BT)]$.  One has to consider all spins of the system, and repeat the "update" over and over again with a large number of times to get thermal equilibrium before calculating statistical thermal averages of physical quantities such as energy, specific heat, magnetization, susceptibility, ...

We need however more sophisticated methods if we wish to calculate critical exponents or to detect  a first-order phase transition.  For calculation of critical exponents, histogram techniques \cite{Ferrenberg,Ferrenberg2} are very precise: comparison with exact results shows an agreement often up to 3rd or 4th digit.  To detect very weak first-order transitions, the Wang-Landau technique \cite{Wang-Landau} combined with the finite-size scaling theory \cite{Barber}  is very efficient. We have used this technique to put an end to a 20-year-old controversy on the nature of the phase transition in Heisenberg and XY frustrated stacked triangular antiferromagnets \cite{Ngo2008b,Ngo2008a}.

\begin{figure}[htb]
\centering
\includegraphics[width=5cm,angle=0]{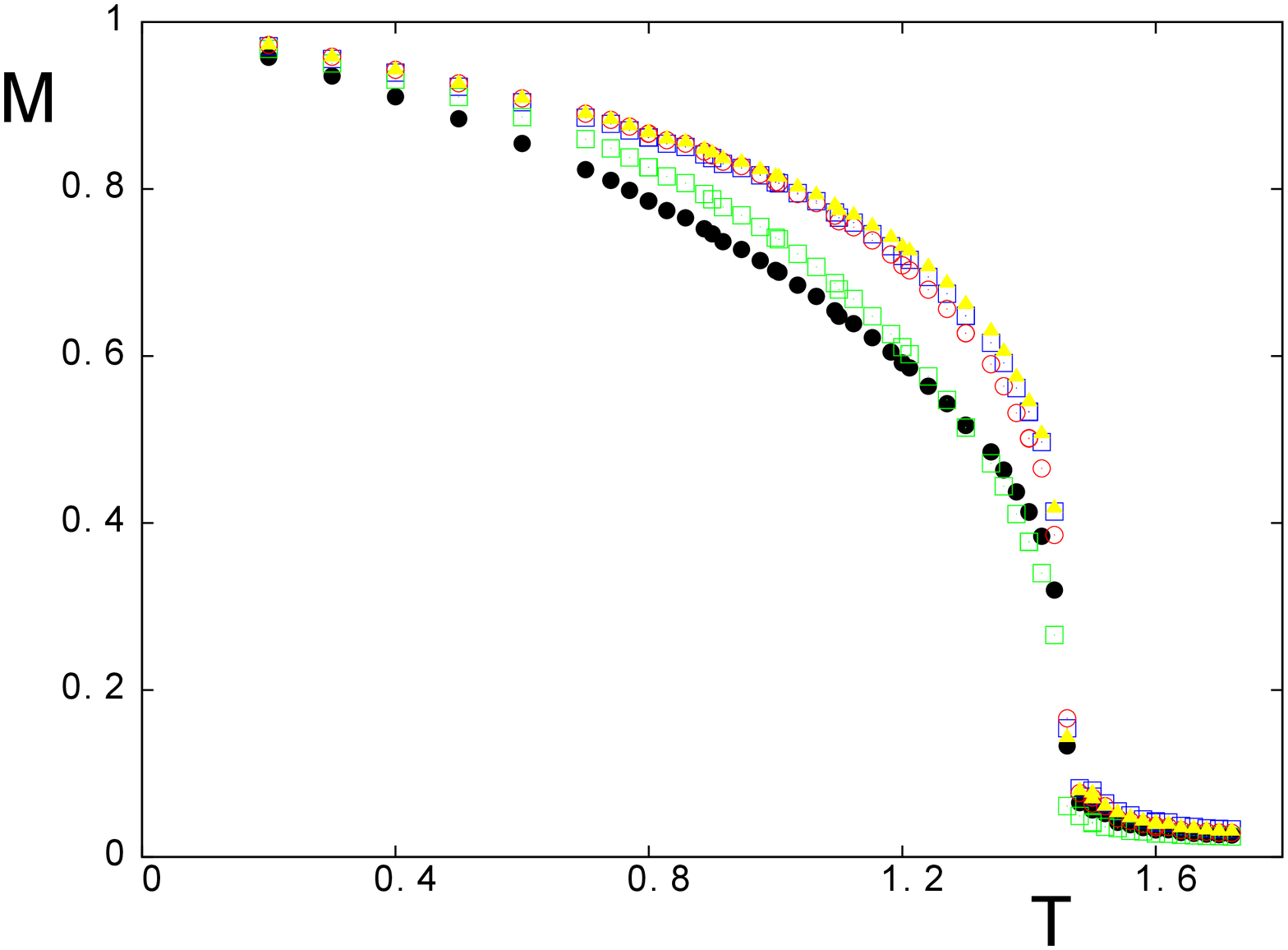}
\includegraphics[width=5cm,angle=0]{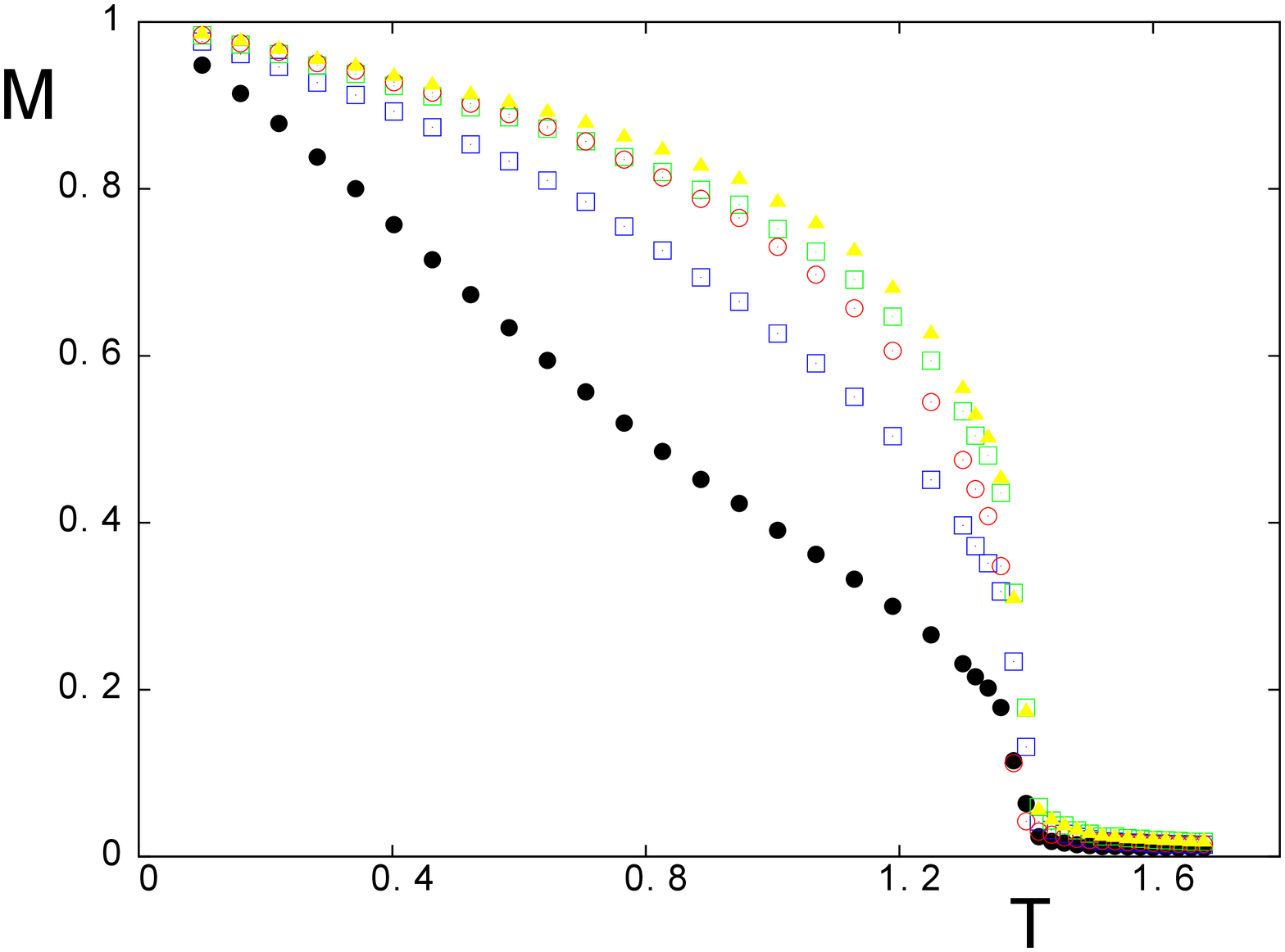}  % .eps
\caption{ (Color online) Monte Carlo results: Layer magnetizations as functions of $T$ for the surface interaction $J_1^s/J_1=1$ (left) and 0.3 (right) with $J_2/J_1=-2$ and $N_z=16$. Black circles, blue void squares, cyan squares and red void circles are for first, second, third and fourth layers, respectively.}\label{surf03MC}
\end{figure}

To illustrate the efficiency of Monte Carlo simulations, let us show in Fig. \ref{surf03MC} the layer magnetizations of the classical counterpart of the body-centered cubic helimagnetic film shown in section \ref{FTF} (figure taken from Ref. \cite{Diep2015}).  Though the surface magnetization is smaller than the magnetizations of interior layers, there is only a single phase transition.
%Fig14

\begin{figure}[htb!]
\centering
\includegraphics[width=5cm,angle=0]{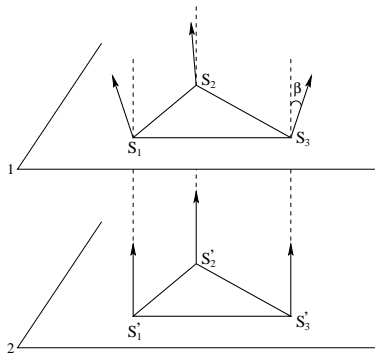}
\caption{Non
collinear surface spin configuration. Angles between spins on
layer $1$ are all equal (noted by $\alpha$), while angles between
vertical spins are $\beta$.} \label{fig:gsstruct}
\end{figure}

To see a surface transition, let us take the case of a frustrated surface of antiferromagnetic triangular lattice coated on a ferromagnetic film of the same lattice structure \cite{NgoSurface}. The in-plane surface interaction is $J_s<0$ and interior interaction is $J>0$. This film has been shown to have a surface spin reconstruction as displayed in Fig. \ref{fig:gsstruct}.
%Fig15

\begin{figure}[hbt!]
\centering
\includegraphics[width=5cm,angle=0]{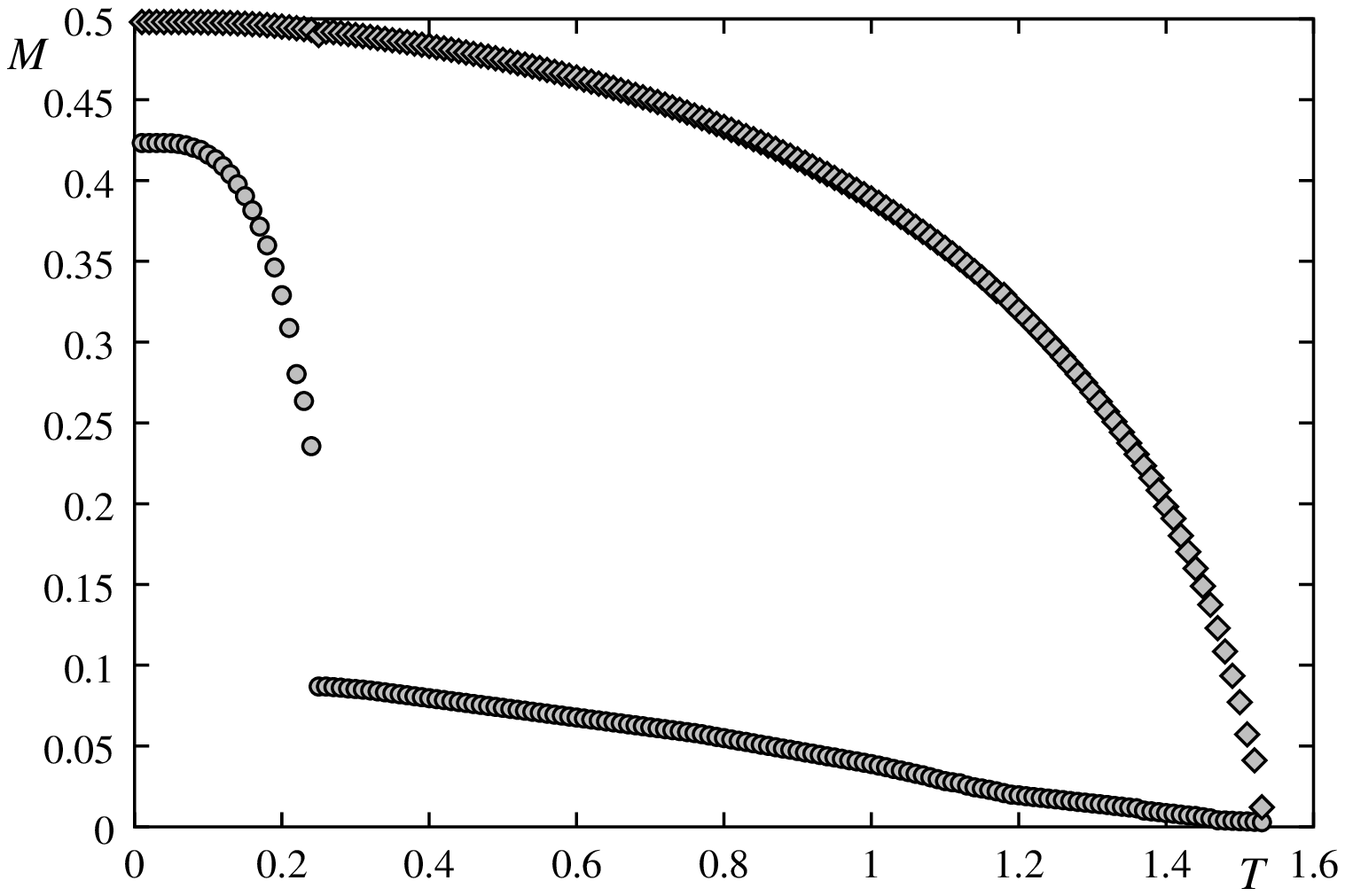}
\includegraphics[width=5cm,angle=0]{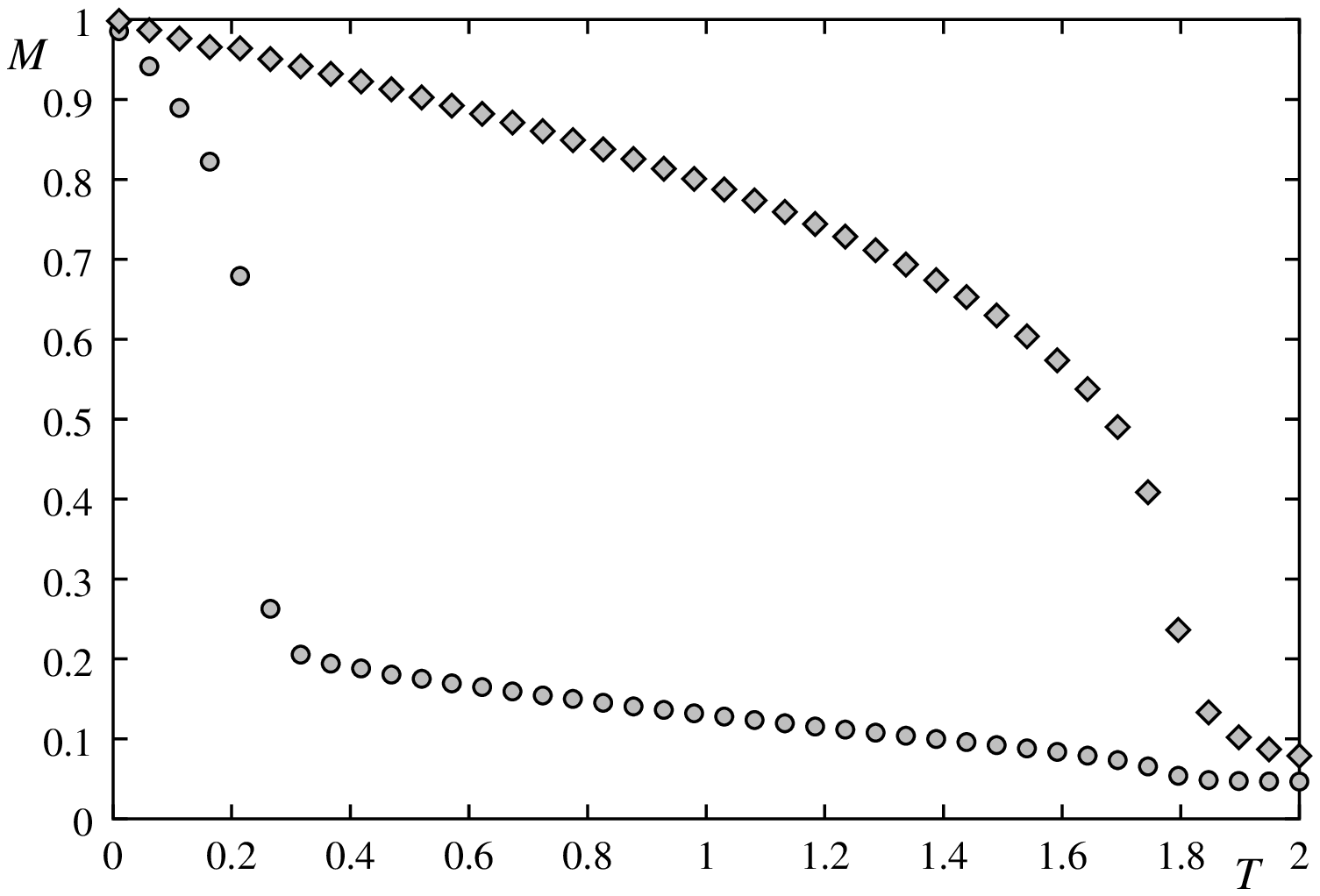}
\caption{Left: First
two layer-magnetizations obtained by the Green's function technique
vs. $T$ for $J_{s} = -0.5J$ with anisotropies $I=-I_s=0.1J$. The surface-layer
magnetization (lower curve) is much smaller than the second-layer
one. Right: Magnetizations of layer 1 (circles) and layer 2
(diamonds) versus temperature $T$ in unit of $J/k_B$. See text for comments.} \label{fig:HGn05Ms}
\end{figure}

We show an example where $J_{s} = -0.5J$ in Fig.
\ref{fig:HGn05Ms}. The left figure is from the Green's function method.
As seen, the surface-layer  magnetization is
much smaller than the second-layer one. In addition there is a
strong spin contraction at $T=0$ for the surface layer. This is
due to quantum fluctuations of the in-plane antiferromagnetic surface
interaction $J_s$.  One sees that the surface becomes disordered
at a temperature $T_1\simeq 0.2557$ while the second layer remains
ordered up to $T_2\simeq 1.522$.   Therefore, the system is
partially disordered for temperatures between $T_1$ and $T_2$.
 This result is very interesting because it confirms again the
existence of the partial disorder in quantum spin systems observed
earlier in the bulk \cite{Quartu1997,Santa1}.  Note that between $T_1$ and $T_2$,
the ordering of the second layer acts as an external field on the
first layer, inducing therefore a small value of its
magnetization.  Results of Monte Carlo simulations of the classical model are shown
on the right of Fig. \ref{fig:HGn05Ms} which have the same features as the quantum case.
%Fig16

%\begin{figure}[thb!]
%\centerline{\epsfig{file=HSn05Ms.eps,width=2.7in}}
%\caption{Magnetizations of layer 1 (circles) and layer 2
%(diamonds) versus temperature $T$ in unit of $J/k_B$ for
%$J_s=-0.5$ with $I=-I_s=0.1$, $L=36$.} \label{fig:HSn05Ms}
%\end{figure}

To close this paragraph we mention that the question of surface criticality has been a long-standing debate. On the one hand, pure theories would tell us that as long as the thickness is finite the correlation in this direction can be renormalized so that the nature of a phase transition in a thin film should be that of the corresponding 2D model. On the other hand, experimental observations and numerical simulations show deviations of critical exponents from 2D universality classes. The reader is referred to Refs. \cite{Pham,Pham1} for discussions on this subject.

\subsection{Surface Reorientation}

In this paragraph, we would like to show the case of thin films where a transition from an in-plane spin configuration to a perpendicular spin configuration is possible at a finite temperature. Such a reorientation occurs when there is a competition between a dipolar interaction which tends to align the spins in the film surface and a perpendicular anisotropy which is known to exist when the film thickness is very small \cite{Heinrich,Zangwill}.   Experimentally, it has been observed in a thin Fe film deposited on a Cu(100) substrate that the perpendicular spin configuration at low temperatures undergoes a transition to a planar spin configuration as the temperature ($T$) is increased \cite{Pappas1,Allenspach,Won,Vaterlaus}. Theoretically, this problem has been studied by many people \cite{Usadel1,Usadel2,Pescia,Santamaria,Hoang}.
Let us consider a 2D surface for simplicity. The case of a film with a small thickness has very similar results \cite{Hoang}. The Hamiltonian includes three parts: a 6-state Potts model ${\cal H}_p$, a dipolar interaction ${\cal H}_d$ and a perpendicular anisotropy ${\cal H}_a$:

\begin{equation}\label{HL}
{\cal H}_p = -\sum_{(i,j)}J_{ij}\delta(\sigma_{i},\sigma_{j} )
\end{equation}
where $\sigma_{i}$ is a variable associated to the lattice site $i$.  $\sigma_{i}$ is equal to 1, 2, 3, 4, 5 and 6 if the spin at that site lies along the $\pm x$,  $\pm y$ and $\pm z$ axes, respectively. $J_{ij}$ is the exchange interaction between NN at $i$ and $j$.  We will assume that (i) $J_{ij}=J_s$ if $i$ and $j$ are on the same film surface (ii) $J_{ij}=J$ otherwise.  For the dipolar interaction, we write

\begin{equation}
{\cal H}_d=D\sum_{(i,j)}\{\frac{\mathbf{S}(\sigma_{i})\cdot \mathbf{S}(\sigma_{j})}{r_{i,j}^3}
-3\frac{[\mathbf{S}(\sigma_{i})\cdot \mathbf r_{i,j}][\mathbf{S}(\sigma_{j})\cdot \mathbf r_{i,j}]}{r_{i,j}^5}\}
\label{dip}
\end{equation}
where $\mathbf r_{i,j}$ is the vector  of modulus $r_{i,j}$  connecting the site  $i$ to the site $j$. One has  $\mathbf r_{i,j}\equiv \mathbf r_j-\mathbf r_i$. $\mathbf{S}(\sigma_{i})$ is a vector of modulus 1 pointing in the $ x$ direction if $\sigma_i=1$, in the $-x$ direction  if $\sigma_i=2$, etc.

The perpendicular anisotropy is
\begin{equation}\label{HA}
{\cal H}_a = -A\sum_{i}s_{z}(i)^2
\end{equation}
where $A$ is a constant.

\begin{figure}
\centering
\includegraphics[width=50mm,angle=0]{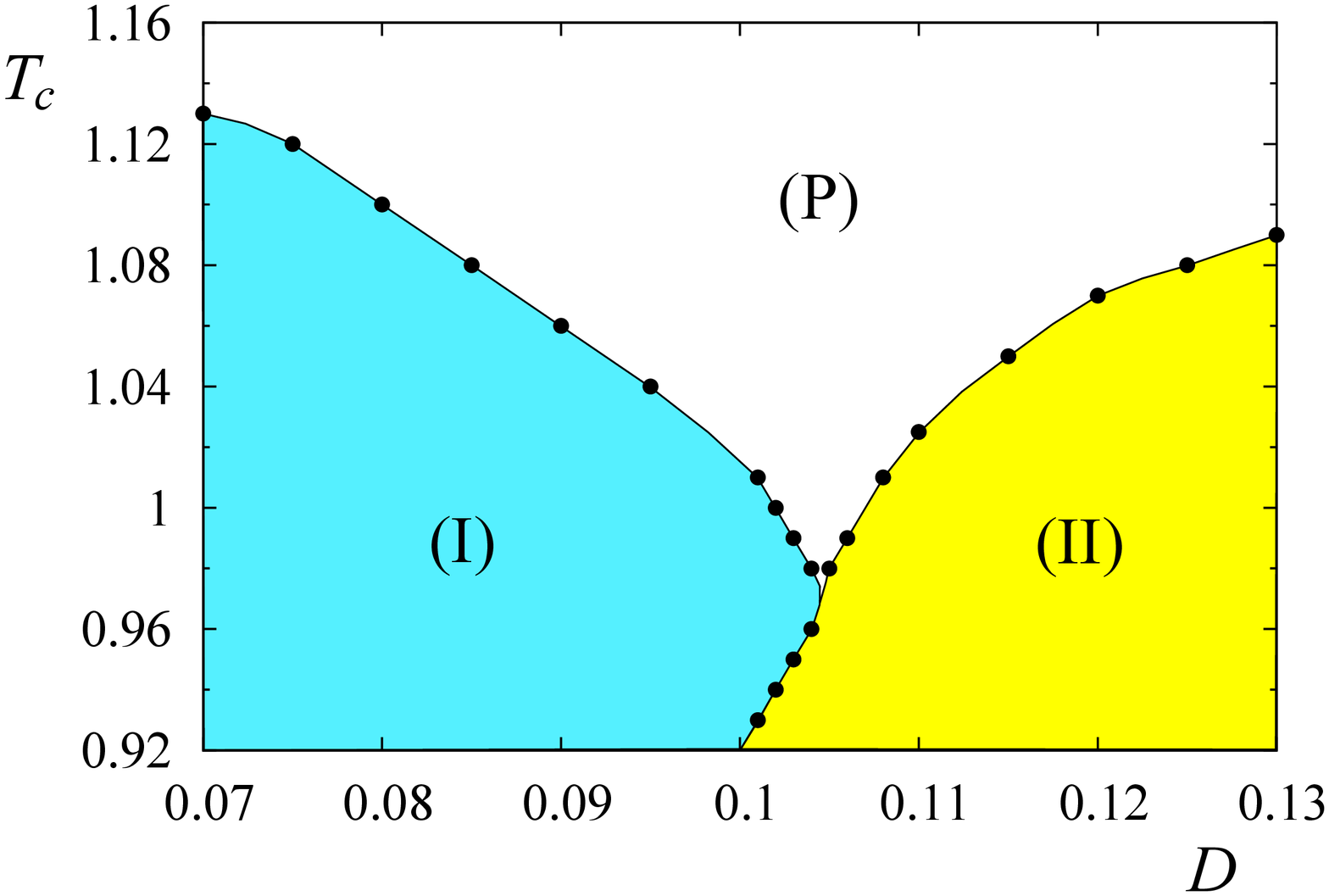}
\includegraphics[width=50mm,angle=0]{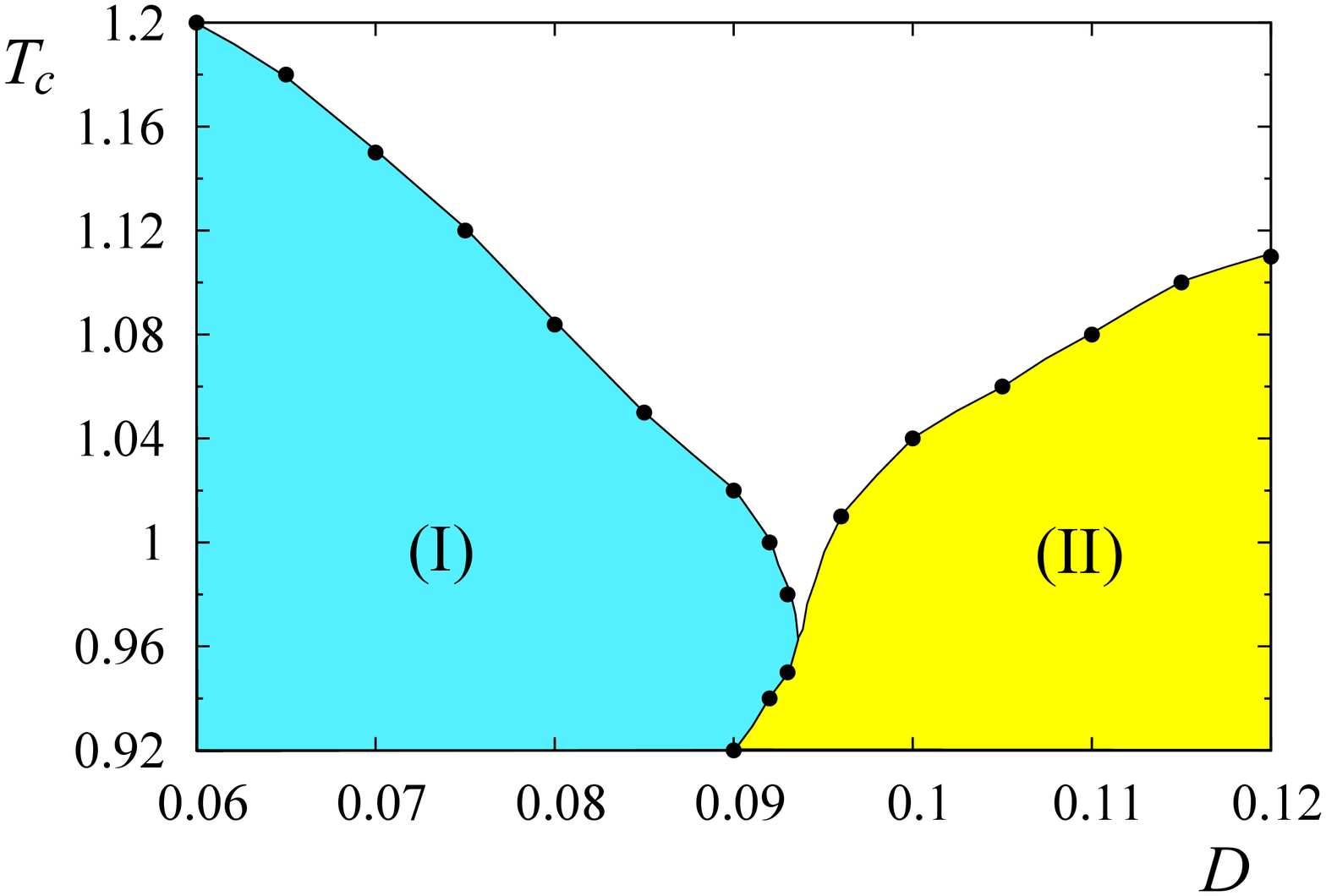}
\caption{(Color online) Phase diagram in 2D.  Transition temperature $T_C$ versus $D$, with $A=0.5$, $J=1$, cutoff distance $r_c=\sqrt{6}$ (left) and $r_c=4$ (right). Phase (I) is the perpendicular spin configuration, phase (II) the in-plane spin configuration and phase (P) the paramagnetic phase. See text for comments. } \label{PD2D}
\end{figure}

Using Monte Carlo simulations, we have established the phase diagram shown in Fig. \ref{PD2D} for two dipolar cutoff distances.  Several remarks are in order: (i) in a small region above $D=0.1$ (left figure) there is a transition from the in-plane to the perpendicular configuration when $T$ increases from 0, (ii) this reorientation is a very strong first-order transition: the energy and magnetization are discontinuous (not shown)
at the transition. Comparing to the Kagom\'e Ising case shown in Fig. \ref{re-fig20}, we do not have a reentrant paramagnetic phase between phases I and II. Instead, we have a first-order transition as also observed near phase frontiers in other systems such as in a frustrated body-centered cubic lattice  \cite{Diep89BCC}.
%Fig17

To conclude this subsection, we mention that competing interactions determine frontiers between phases of different symmetries. Near these frontiers, we have seen many interesting phenomena such as reentrance, disorder lines, reorientation transition, ...  when the temperature increases.

\subsection{Spin Transport in Thin Films}
The total resistivity stem  from different kinds of diffusion processes in a crystal. Each contribution has in general a different temperature dependence.
Let us summarize the most important contributions to the total resistivity $\rho_t(T)$ at low temperatures in the following expression
\begin{equation}\label{rhot}
\rho_t(T)=\rho_0+A_1T^2+A_2T^5+A_3\ln{\frac{\mu}{T}}
\end{equation}
where $A_1$, $A_2$ and $A_3$ are constants. The first term is $T$-independent,
the second term proportional to $T^2$ represents the scattering of itinerant spins at low $T$ by lattice spin-waves. Note that the resistivity caused by a Fermi liquid is also proportional to $T^2$.   The $T^5$ term corresponds to a low-$T$ resistivity in metals.  This is due to the scattering of itinerant electrons by phonons.  At high $T$, metals however show a linear-$T$ dependence.  The logarithm term is the resistivity due to the quantum Kondo effect caused by a magnetic impurity at very low $T$.

We are interested here in the spin resistivity $\rho$ of magnetic materials.
We have developed an algorithm which allows us to calculate the spin resistivity in various magnetically ordered systems \cite{Akabli1,Akabli2,Akabli3,Akabli4,Hoang1,Magnin1,Magnin2,Magnin1,Magnin2}, in particular in thin films.  Unlike the charge conductivity, studies of spin transport have been regular  but not intensive until recently. The situation changes   when the electron spin begins to play a central role in spin electronics, in particular with the discovery of the colossal magnetoresistance \cite{Fert,Grunberg}.

The main mechanism which governs the spin transport is the interaction between itinerant electron spins and localized spins of the lattice ions ($s-d$ model).  The spin-spin correlation has been shown to be responsible for the behavior of the spin resistivity \cite{DeGennes,Fisher,Kasuya}. Calculations were mostly done by mean-field approximation (see references in \cite{Kataoka}). Our works mentioned above were the first to use intensive Monte Carlo simulations for investigating the spin transport. We also used a combination of the Boltzmann equation \cite{DiepTM,Cercignani} and numerical data obtained by simulations \cite{Akabli3,Akabli4}.
The Hamiltonian includes three main terms: interaction between lattice spins $\mathcal{H}_l $,  interaction between itinerant spins and lattice spins $\mathcal{H}_r$, and  interaction between itinerant spins $\mathcal{H}_m$. We suppose

\begin{eqnarray}
\mathcal{H}_l & = & -\sum_{(i,j)}J_{i,j}\mathbf{S}_{i}\cdot\mathbf{S}_{j}\label{HamilR}
\end{eqnarray}
where $\mathbf {S}_i$ is the spin localized at lattice site
$i$ of Ising, XY or Heisenberg model, $J_{i,j}$ the exchange integral
between the spin pair $\mathbf S_i$ and $\mathbf S_j$ which is not limited to the interaction between nearest-neighbors (NN).  For $\mathcal{H}_r$ we write

\begin{eqnarray}
 \mathcal{H}_r & = & -\sum_{i,j}I_{i,j}\mathbf{\sigma}_i \cdot \mathbf{S}_j\label{I}
 \end{eqnarray}
 where $\mathbf{\sigma}_i$ is the spin of the i-th itinerant electron and $I_{i,j}$ denotes the interaction which depends on the distance between  electron $i$ and  spin $\mathbf{S}_j$ at  lattice site $j$. For simplicity, we suppose the following interaction expression
  \begin{eqnarray} \label{IJ}
  I_{i,j} & = & I_{0}e^{-\alpha r_{ij}}
  \end{eqnarray}
  where $r_{ij}=|\vec{r}_i-\vec{r}_j|$,  $I_0$ and $\alpha$ are  constants.
  We use a cut-off distance $D_1$ for the above interaction.  Finally, for $\mathcal{H}_m$, we use
   \begin{eqnarray}
  \mathcal{H}_m & = & -\sum_{i,j}K_{i,j}\mathbf{\sigma}_i \cdot \mathbf{\sigma}_j\\
  \mbox{where}\ \ K_{i,j} & = & K_{0}e^{-\beta r_{ij}}\label{K}
  \end{eqnarray}
  with $K_{i,j}$ being the interaction  between electrons $i$ and $j$.
  The system is under an electrical field $\vec \epsilon$ which creates an electron current in one direction. In addition we include also a chemical potential which keeps electrons uniformly distributed in the system. Simulations have been carried out with the above Hamiltonian.  The reader is referred to the original papers mentioned above for  technical details. As expected, the spin resistivity reflects the spin-spin correlation of the system: $\rho$ has the form of the magnetic susceptibility, namely it shows a peak at the phase transition.  It is noted however that unlike the susceptibility which diverges at the transition, the spin resistivity is finite at $T_c$ due to the fact that only short-range correlations can affect the resistivity (see arguments given in \cite{Fisher}). Moreover, it is known that near the phase transition the system is in the critical-slowing-down regime. Therefore, care should be taken while simulating in the transition region. This point has been considered in our simulations by introducing the relaxation time \cite{Magnin3}.

\begin{figure}[h!]
 \centering
  \includegraphics[width=45mm,angle=0]{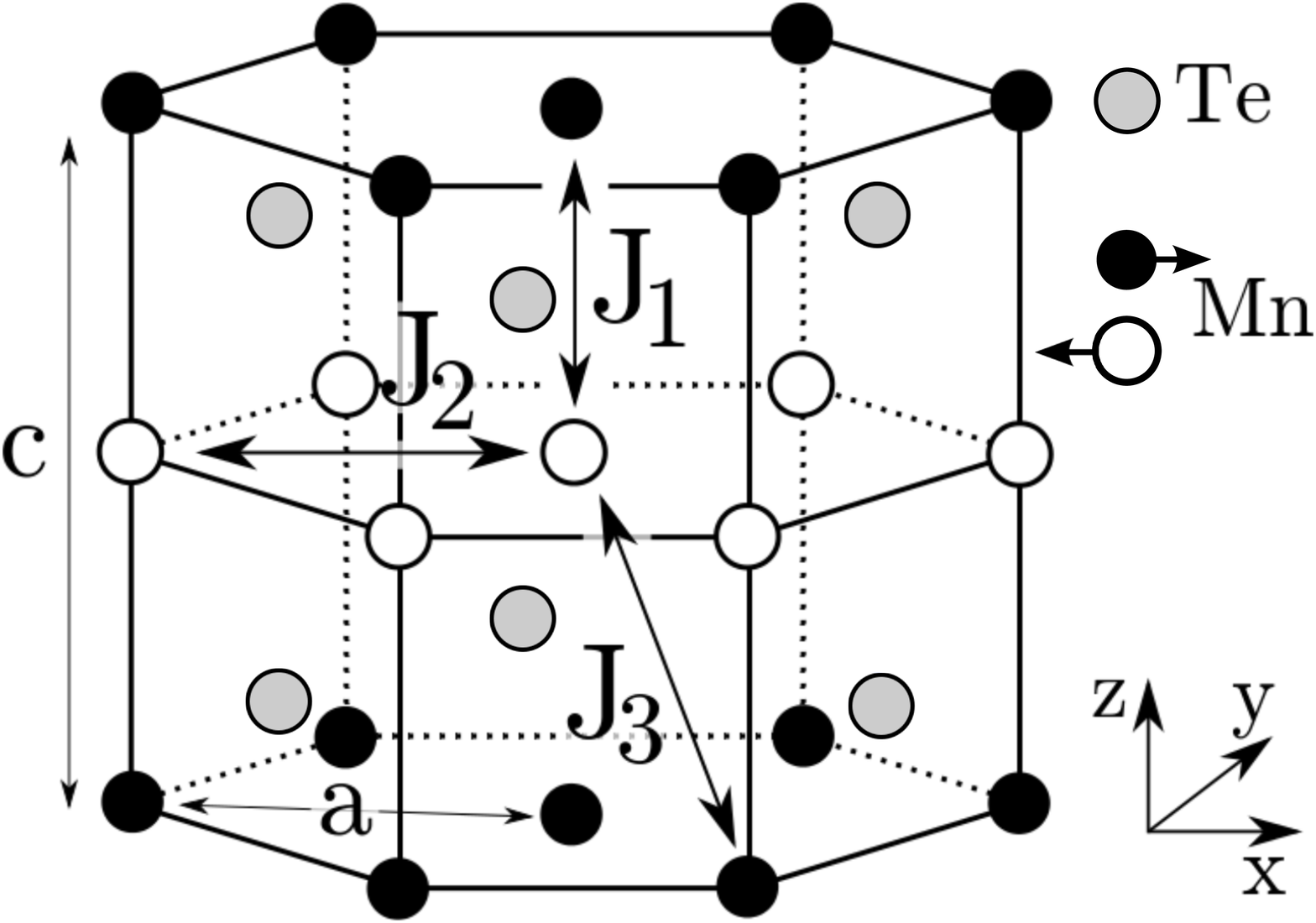}
  \includegraphics[width=65mm,angle=0]{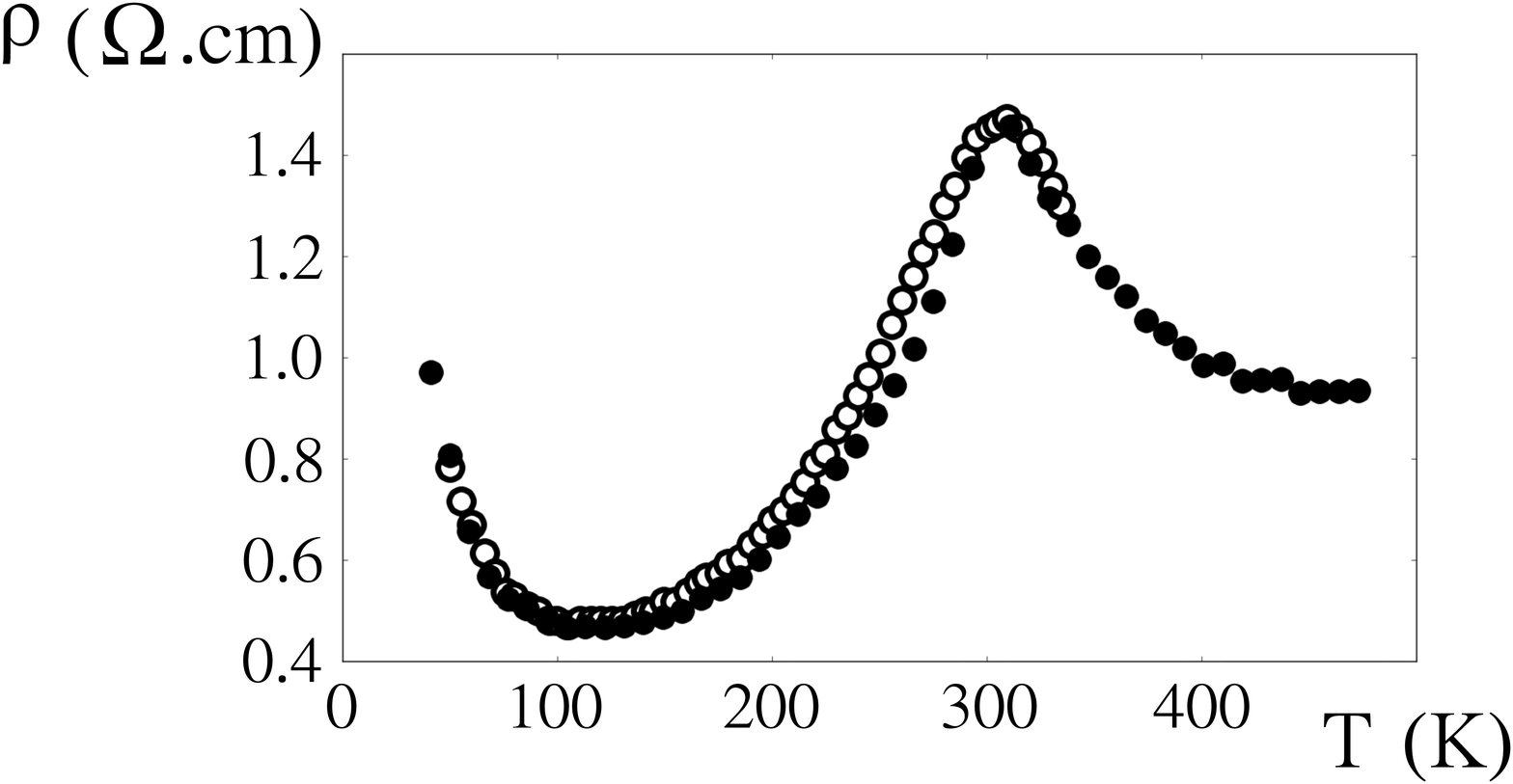}
 \caption{Left: Structure of MnTe of  NiAs type. Antiparallel spins are shown by black and white circles.  NN interaction is marked by $J_1$, NNN interaction by $J_2$, and third NN one by $J_3$.
 Right: $\rho$ versus $T$. Black circles are from Monte Carlo simulation, white circles are experimental data taken from He et al.\cite{He}. The parameters used in the simulation are taken from \cite{Hennion2}: $J1=-21.5$K, $J_2=2.55$ K, $J_3=-9$ K, $D_a=0.12$ K (anisotropy), $D_1=a=4.148\AA$, and $I_0=2$ K, $\epsilon = 2*10^5$ V/m. } \label{MnTe}
\end{figure}

  To illustrate the efficiency of Monte Carlo simulations, we show in Fig. \ref{MnTe} the excellent agreement of our simulation \cite{Magnin4} and experiments performed on MnTe \cite{He}.  The interactions and the crystalline parameters were taken from Ref. \cite{Hennion2}.
  %Fig18

When a film has a surface phase transition at a low temperature in addition to the transition of the bulk at a higher temperature, one observes two peaks in the spin resistivity as shown in Ref. \cite{Akabli3}.

%  \begin{figure}
%%\includegraphics[width=80mm,height=40mm]{fig1.jpg}
%\centering \includegraphics[width=40mm,angle=-90]{FIG5sup-new.ps}
%\centering \includegraphics[width=40mm,angle=0]{FIG6.ps}
%\caption{Left: Magnetization  versus $T$ in the case where
%the system is made of three films: the first and the third have 5
%layers with a weaker interaction $J_s$, while the middle has 4
%layers with interaction $J=1$. We take $J_s=0.2 J$.  Black
%triangles: magnetization of the surface films, stars:
%magnetization of the middle film, void circles: total
%magnetization. Right: Resistivity $\rho$ in arbitrary unit versus  $T$. See text for
%comments.}\label{surf1}
%\end{figure}

%\section{Interpretation of Experimental Results: Difficulties and Uncertainties}\label{TversusE}

\section{Conclusion}\label{Conclu}
To conclude this review let us discuss on the relation between theories and experiments, in particular on the difficulties encountered when one is confronted, on the one hand, with simplified theoretical pictures and hypotheses and, on the other hand, with insufficient experimental knowledge of what is really inside the material. We would like to emphasize on the importance of a sufficient theoretical background to understand experimental data measured on systems which are more complicated, less perfect than models used to describe them.
Real systems have always impurities, defects, disorder, domains, ... However, as long as these imperfections are at extremely small amounts, they will not affect observed macroscopic quantities:  theory tells us that each observable is a result from a statistical average over all microscopic states and over the space occupied by the material. Such an averaging will erase away rare events leaving only common characteristics of the system.  Essential aspects can be thus understood from simple models if one includes correct ingredients based on physical arguments while constructing the model.

One of the striking points shown above is the fact that without sophisticated calculations, we cannot discover hidden secrets of the nature such as the existence of disorder lines with and without dimension reduction, the extremely narrow reentrant region between two ordered phases, the coexistence of order and disorder of a system at equilibrium etc.  These effects are from the competition between various interactions which are unavoidable in real materials.  These interactions determine the boundaries between various phases of different symmetries in the space of physical parameters. Crossing a boundary the system will change its symmetry. Theory tells us that if the symmetry of one phase is not a subgroup of the other then the transition should be of first order or the two phases should be separated by a narrow reentrant phase. Without such a knowledge, we may overlook such fine effects while examining experimental data.

We have used frustrated thin films to illustrate various effects due to a combination of frustration and surface magnetism. We have seen that to understand when and why surface magnetization is small with respect to the bulk one we have to go through a microscopic mechanism to recognize that low-lying localized surface spin-wave modes when integrated in the calculation of the magnetization will indeed mathematically lower its value. Common effects observed in thin films such as surface reconstruction and surface disordering can be theoretically explained.

We should emphasize on the importance of a combination of Monte Carlo simulation and theory. We have seen for example that we can determine the exact transition temperature in the Kagom\'e model but the nature of the ordering is understood only with Monte Carlo simulation. This is not the only example: as theorists we need also hints and checks while constructing a theory, and as a simulator we need to understand what comes out from the computer since we often loose the track between inputs and outputs in a simulation. In this respect, a combination of numerical and theoretical methods is unavoidable.

\vspace{1cm}

%\Acknowledgments
I am grateful to my numerous former and present doctorate students with whom I shared uncountable wonderful moments of scientific discussions and from whom I draw my source of energy and joy.




%\section*{References}

\end{document}